\documentclass[aps,prb,superscriptaddress,showpacs,floatfix]{revtex4}
\usepackage{graphicx}
\usepackage{amsmath}
\usepackage{amssymb}
\usepackage{dcolumn}

\begin{document}

\title{Ground-state densities and pair correlation functions in parabolic quantum dots}
\author{M. Gattobigio}
\email{gatto@sns.it}
\affiliation{NEST-INFM and Classe di Scienze, Scuola Normale Superiore, I-56126 Pisa, Italy}
\author{P. Capuzzi}
\email{capuzzi@sns.it}
\affiliation{NEST-INFM and Classe di Scienze, Scuola Normale Superiore, I-56126 Pisa, Italy}
\author{M. Polini}
\email{m.polini@sns.it}
\affiliation{NEST-INFM and Classe di Scienze, Scuola Normale Superiore, I-56126 Pisa, Italy}
\author{R. Asgari}
\email{asgari@theory.ipm.ac.ir}
\affiliation{Institute for Studies in Theoretical Physics and Mathematics, Tehran 19395-5531, Iran}
\affiliation{NEST-INFM and Classe di Scienze, Scuola Normale Superiore, I-56126 Pisa, Italy}
\author{M. P. Tosi}
\email{tosim@sns.it}
\affiliation{NEST-INFM and Classe di Scienze, Scuola Normale Superiore, I-56126 Pisa, Italy}

\begin{abstract}
We present an extensive comparative study of ground-state densities and pair distribution functions for 
electrons confined in two-dimensional parabolic quantum dots 
over a broad range of coupling strength and electron number. We first use spin-density-functional theory 
to determine spin densities that are compared with Diffusion Monte Carlo (DMC) data. 
This accurate knowledge of one-body properties is then used to construct and 
test a local approximation for the electron-pair correlations. 
We find very satisfactory agreement between this local scheme and the 
available DMC data, and provide a detailed picture of two-body correlations in 
a coupling-strength regime preceding the formation of Wigner-like electron ordering.
\end{abstract}
\maketitle

\section{Introduction}
For a number of years there has been a growing interest in studying finite quantum systems under external confinement, such as ultracold atomic or molecular gases inside magnetic or optical traps~\cite{bec} and electrons in metallic clusters~\cite{deHeer} or quantum dots~\cite{general_qd}. The confinement introduces a new length scale and induces novel physical behaviors relative to the corresponding infinitely extended model system. In particular, in a quantum dot the properties of a homogeneous electron gas are profoundly modified by the emergence of effects that are commonly associated with electrons in atoms. A well-known example is the presence of a shell structure in the energy to add electrons to a quantum dot~\cite{addition_energy}.

With regard to the spatial structure of the electronic system, the analogue of two-dimensional (2D) Wigner crystallization has been shown in a path-integral Monte Carlo study~\cite{filinov_2001} to occur in two distinct stages inside a circularly symmetric parabolic quantum dot. Radial ordering of the electrons into shells occurs first and is followed by orientational ordering through freezing of intershell rotations. Short-range order in the electronic structure at lower coupling strength is described by the pair distribution function $g({\bf r},{\bf r}')$ giving the spin-averaged probability of finding two electrons at positions ${\bf r}$ and ${\bf r}'$. Some properties of this function and of its extension to describe spin-resolved pair correlations have been reported for a circular quantum dot in a DMC study by Pederiva {\it et al.}~\cite{pederiva_2000}. In the macroscopic limit $g({\bf r},{\bf r}')$ reduces to a function of the relative distance $|{\bf r}-{\bf r}'|$ of an electron pair and describes the liquid-like short-range order in the homogenous electron gas.

The main purpose of this work is to present an approximate theoretical treatment of the electron-pair correlations in quantum dots at such values of the coupling strength and of the electron number. We use spin-density-functional theory~\cite{dreizler_book} (SDFT) in local spin density approximations and test their accuracy against the available DMC data of Ref.~\onlinecite{pederiva_2000} for both one-body and two-body structural properties. In brief, the paper is organized as follows. Section~\ref{sect:bh} summarizes the SDFT procedure for the sake of completeness and in order to set the stage for the later sections of the paper. Our results are presented and discussed in Section~\ref{sect:ed} for the one-body spin densities and in Section~\ref{sect:paircorrelation} for the pair correlation functions. An Appendix emphasizes the distinction between the short-range order of present interest and the broken-symmetry states that are met in unrestricted Hartree-Fock (HF) calculations~\cite{yannouleas_1999} on electrons in quantum dots. Finally, a brief summary of our main conclusions is given in Section~\ref{sect:conclusions}.
\label{sect:intro}

\section{Theoretical Approach}
\label{sect:bh}

We consider $N$ interacting electrons of band mass $m$, 
confined in a strictly $2D$ parabolic quantum dot (QD). 
The real-space Hamiltonian is
\begin{eqnarray}\label{eq:hamiltonian}
{\hat {\mathcal H}}_{\rm QD}&=&-\frac{\hbar^2}{2m}
\sum_{\sigma}\int \,d^2{\bf r} \, {\hat \psi}^{\dagger}_{\sigma}({\bf r})
\,\nabla^2\,{\hat \psi}_{\sigma}({\bf r})+
\sum_{\sigma}\int \,d^2{\bf r} \,V_{\rm ext}({\bf r}) {\hat \psi}^{\dagger}_{\sigma}({\bf r})
{\hat \psi}_{\sigma}({\bf r})
\nonumber \\
&+&\frac{1}{2}\sum_{\sigma, \sigma'}\int
\,d^2{\bf r} \int \,d^2{\bf r'} 
\,{\hat \psi}^{\dagger}_{\sigma}({\bf r})
{\hat \psi}^{\dagger}_{\sigma'}({\bf r'})
v(|{\bf r}-{\bf r'}|)
{\hat \psi}_{\sigma'}({\bf r'}){\hat \psi}_{\sigma}({\bf r})\,.
\end{eqnarray}
Here ${\hat \psi}_{\sigma}({\bf r})$ and ${\hat \psi}^{\dagger}_{\sigma}({\bf r})$ are 
Schr\"{o}dinger field operators obeying canonical anticommutation
relations, $V_{\rm ext}({\bf r})=m\omega_0^2{\bf r}^2/2$ is the external confining potential and $v(r)=e^2/(\kappa r)$ is the interparticle Coulomb potential, $\kappa$ being the dielectric constant of the material. The Hamiltonian (\ref{eq:hamiltonian}) commutes with the $z$-component $S_z$ 
of the total spin and therefore $S_z$ is a good quantum number.

We introduce the spin density $n_{\sigma}({\bf r})=\langle{\hat \psi}^{\dagger}_{\sigma}({\bf r}){\hat \psi}_{\sigma}({\bf r})\rangle$, the density matrix $\rho_{\sigma\sigma'}({\bf r},{\bf r}')=\langle{\hat \psi}^{\dagger}_{\sigma}({\bf r}){\hat \psi}_{\sigma'}({\bf r}')\rangle$, and the two-body density $n^{(2)}_{\sigma\sigma'}({\bf r},{\bf r}')=\langle{\hat \psi}^{\dagger}_{\sigma}({\bf r}){\hat \psi}^{\dagger}_{\sigma'}({\bf r'})
{\hat \psi}_{\sigma'}({\bf r'}){\hat \psi}_{\sigma}({\bf r})\rangle=n_{\sigma}({\bf r})n_{\sigma'}({\bf r}')
g_{\sigma\sigma'}({\bf r},{\bf r}')$, $g_{\sigma\sigma'}({\bf r},{\bf r}')$ being the spin-resolved 
pair distribution function (PDF). The calculations will be carried out at fixed $N=N_\uparrow+N_\downarrow$ and $S_z=(N_\uparrow-N_\downarrow)/2$, where $N_\sigma=\int
d^2{\bf r}\,n_\sigma({\bf r})$ is the total number of electrons with spin $\sigma=\uparrow, \downarrow$.

Choosing as unit of length the harmonic-oscillator length $\ell_0=\sqrt{\hbar/(m\omega_0)}$ and 
as unit of energy the harmonic-oscillator quantum $\hbar\omega_0$, the QD Hamiltonian can be shown to be governed by 
the dimensionless parameter
\begin{equation}\label{eq:lambda}
\lambda=\frac{e^2/(\kappa \ell_0)}{\hbar\,\omega_0}=\frac{\ell_0}{a^\star_B}\,,
\end{equation}
where $a^\star_B=\kappa\hbar^2/(me^2)$ is the effective Bohr radius. The physical properties of the electron assembly are thus functions of the quantities $\lambda$, $N$, and $S_z$. We will choose parameters that are suitable for a $2D$ electron gas (EG) confined in a GaAs quantum well, {\it i.e.} $m=0.067$ bare electron masses and $\kappa=12.4$. With this choice $a^\star_B\approx 9.8\,{\rm nm}$ and the effective Hartree energy is $e^2/(\kappa a^\star_B)\approx 11.9\,{\rm meV}$.

\subsection{Spin-density-functional theory in the Kohn-Sham scheme}
\label{sect:sdft_ks}

Within the Kohn-Sham (KS) version of SDFT the calculation of the equilibrium densities $n_\sigma({\bf r})$ is recast into the solution of a set of Schr\"{o}dinger-like equations for the single-particle orbitals $\phi^{\rm KS}_{i,\sigma}({\bf r})$,
\begin{equation}\label{eq:ks}
\left[-\frac{\hbar^2}{2m}\nabla^2+V_{\rm ext}({\bf r})+v_{\rm H}({\bf r};[n_\sigma])+v^{\sigma}_{\rm xc}({\bf r};[n_\sigma])\right]\phi^{\rm KS}_{i,\sigma}({\bf r})=\varepsilon_{i,\sigma}\phi^{\rm KS}_{i,\sigma}({\bf r})\,.
\end{equation}
Here 
$
v_{\rm H}({\bf r};[n_\sigma])=\sum_{\sigma}\int d^2{\bf r}'v(|{\bf r}-{\bf r}'|)n_{\sigma}({\bf r}')
$ 
is the classical Hartree potential and 
$
v^{\sigma}_{\rm xc}({\bf r};[n_\sigma])=\delta E_{\rm xc}[n_\sigma]/\delta n_{\sigma}({\bf r})
$
is the spin-dependent exchange-correlation (xc) potential defined as the functional derivative of the 
xc energy functional $E_{\rm xc}[n_\sigma]$. The approximation that we have employed for $v^{\sigma}_{\rm xc}({\bf r};[n_\sigma])$ is discussed below.

The KS mapping guarantees that the 
equilibrium spin densities can be built from the KS orbitals in a single-particle fashion, 
\begin{equation}\label{eq:closure}
n_{\sigma}({\bf r})=\sum_{i\,({\rm occ})}\left|\phi^{\rm KS}_{i,\sigma}({\bf r})\right|^2\,,
\end{equation}
where the sum runs over all occupied states. This equation also provides a self-consistent closure for the KS equations (\ref{eq:ks}). Once these equations have been solved, the ground-state energy ${\mathcal E}$ of the system 
is obtained from ${\mathcal E}=T_{\rm s}+E_{\rm H}[n_{\sigma}]+V_{\rm ext}[n_{\sigma}]+E_{\rm xc}[n_{\sigma}]$, where $T_{\rm s}$ is the kinetic energy of the auxiliary noninteracting electron system, 
$E_{\rm H}[n_{\sigma}]$ is the Hartree potential energy, and $V_{\rm ext}[n_{\sigma}]$ is the contribution 
from the external potential. 

\subsection{Adiabatic connection and approximate xc potential}
\label{sect:sdft_acf}

An implicit expression for the xc energy functional, which highlights the importance of the PDF, 
is the adiabatic connection formula~\cite{perdew_1975}. This reads
\begin{equation}\label{eq:adiabatic_connection}
E_{\rm xc}[n_{\sigma}]=\frac{1}{2}\sum_{\sigma, \sigma'}\int d^2{\bf r}\int d^2{\bf r}'\,
v(|{\bf r}-{\bf r}'|)n_{\sigma}({\bf r})n_{\sigma'}({\bf r}')\,[{\bar g}_{\sigma\sigma'}({\bf r},{\bf r}';[n_{\sigma}])-1]\,,
\end{equation}
where
\begin{equation}\label{eq:gbar}
{\bar g}_{\sigma\sigma'}({\bf r},{\bf r}';[n_{\sigma}])=\int_0^1 d\xi\, g^{(\xi)}_{\sigma\sigma'}({\bf r},{\bf r}';[n_{\sigma}])
\end{equation}
is the coupling-constant averaged PDF. Here 
$g^{(\xi)}_{\sigma\sigma'}$ is the PDF for a system with interactions $v_\xi(r)=e^2\xi/(\kappa r)$ and fixed 
($\xi$-independent) equilibrium densities $n_{\sigma}({\bf r})$. This function depends on $n_{\sigma}({\bf r})$ as a result of the Hohenberg-Kohn theorem. 

The local spin density approximation~\cite{gunnarsson_1989} (LSDA) for $E_{\rm xc}[n_{\sigma}]$ reads
\begin{equation}\label{eq:lda}
E^{\rm LSDA}_{\rm xc}[n_{\sigma}]=\int d^2{\bf r}\,n({\bf r})\left.\varepsilon^{\rm hom}_{\rm xc}(n,\zeta)\right|_{n\rightarrow n({\bf r}),\zeta \rightarrow \zeta({\bf r})}\,,
\end{equation}
where $\varepsilon^{\rm hom}_{\rm xc}(n,\zeta)$ is the xc energy per particle of a homogeneous $2D$ electron gas (EG) as a function of the total particle density $n=\sum_\sigma n_\sigma$ and of the spin-polarization $\zeta=\sum_\sigma\sigma n_\sigma/n$. Accurate results for $\varepsilon^{\rm hom}_{\rm xc}(n,\zeta)$ have been obtained in Quantum Monte Carlo simulations by Attaccalite {\it et al.}~\cite{attaccalite_2002} with special attention to its dependence on $\zeta$.

The adiabatic connection formula allows one to interpret the LSDA as an approximate choice  for ${\bar g}_{\sigma\sigma'}({\bf r},{\bf r}';[n_\sigma])$. One readily obtains Eq.~(\ref{eq:lda}) by 
approximating the xc energy density from the exact expression in Eq.~(\ref{eq:adiabatic_connection}) 
with $n\,\varepsilon^{\rm hom}_{\rm xc}(n,\zeta)$ taken at the local density $n({\bf r})$ and at the local spin-polarization $\zeta({\bf r})$. The xc energy of the EG is given by 
\begin{equation}\label{eq:2degedensity}
\varepsilon^{\rm hom}_{\rm xc}(n,\zeta)=\pi\sum_{\sigma,\sigma'}\frac{n_{\sigma}n_{\sigma'}}{n}
\int_0^{+\infty} rdr\,v(r)\,[{\bar g}^{\rm hom}_{\sigma\sigma'}(r;n,\zeta)-1]\,,
\end{equation}
where ${\bar g}^{\rm hom}_{\sigma\sigma'}(r;n,\zeta)$ is the coupling-constant averaged PDF in the $2D$ EG. 
Thus within the LSDA the exact functional $g_{\sigma\sigma'}({\bf r},{\bf r}';[n_\sigma])$ is approximated as
\begin{equation}\label{eq:glsda}
g_{\sigma\sigma'}({\bf r},{\bf r}';[n_\sigma])\simeq \left. g^{\rm hom}_{\sigma\sigma'}(|{\bf r}-{\bf r}'|;n,\zeta)\right|_{n\rightarrow n({\bf r}),\zeta \rightarrow \zeta({\bf r})}\equiv 
\left.g_{\sigma\sigma'}({\bf r},{\bf r}')\right|_{\rm LSDA}\,.
\end{equation}
Analytical representations of accurate Quantum Monte Carlo data 
for the spin-averaged PDF of the homogeneous $2D$ EG are available in the literature~\cite{gori_giorgi_2004} 
and provide a convenient input for our work.

We recall at this point for later use that the exact PDF satisfies the so-called central sum rule
\begin{equation}\label{eq:central_sum_rule}
\int d^2{\bf r}'n_{\sigma}({\bf r}')\left[g_{\sigma'\sigma''}({\bf r},{\bf r}')-1\right]
=-\delta_{\sigma\sigma'}\delta_{\sigma\sigma''}\,,
\end{equation}
in addition to the symmetry property 
$
g_{\sigma\sigma'}({\bf r},{\bf r}')=g_{\sigma'\sigma}({\bf r}',{\bf r})
$
and to the asymptotic result $\lim_{|{\bf r}-{\bf r}'|\rightarrow +\infty}g_{\sigma\sigma'}({\bf r},{\bf r}')=1$.

\subsection{Fock-Darwin basis}
\label{sect:sdft_fd}

In the numerical solution of the KS equations we have adopted a standard 
procedure involving projection of Eq.~(\ref{eq:ks}) on the Fock-Darwin (FD) basis corresponding to the (complete and orthonormal) set of eigenfunctions of the $2D$ isotropic harmonic oscillator. These are the product of the eigenstates of the angular momentum ${\hat L}_z=-i\hbar\partial_\theta$ and of the radial functions $R_{n,M}(r)$,
$
\varphi_{n,M}({\bf r})=(2\pi)^{-1/2}\exp{(iM\theta)}R_{n,M}(r)
$.
The quantum numbers $n$ and $M$ represent the number of nodes of $R_{n,M}(r)$ 
and the angular momentum $M\hbar$ carried by the state.
The radial wave functions are expressed through the generalized Laguerre polynomials~\cite{abram}. 

The projection of Eq.~(\ref{eq:ks}) onto the FD basis is straightforward. Decomposition of the KS orbitals, $\phi^{\rm KS}_{i,\sigma}({\bf r})=\sum_{\alpha}C^{i,\sigma}_\alpha\varphi_{\alpha}({\bf r})$ where $\alpha$ stands for the pair $\{n_{\alpha},M_{\alpha}\}$, leads to a matrix eigenvalue problem~\cite{szabo_1989} for the coefficients $C^{i,\sigma}_\alpha$,
\begin{equation}\label{eq:ks_p}
\sum_{\beta}\langle\alpha|{\mathcal H}_{\rm KS}|\beta\rangle C^{i,\sigma}_\beta=\varepsilon_{i,\sigma}C^{i,\sigma}_\alpha\,.
\end{equation}
Here ${\mathcal H}_{\rm KS}$ is the effective KS Hamiltonian in Eq.~(\ref{eq:ks}).
The equilibrium densities in Eq.~(\ref{eq:closure}) take the form $n_{\sigma}({\bf r})=\sum_{\alpha,\beta}(\sum_i[C^{i,\sigma}_\alpha]^*C^{i,\sigma}_\beta)
\varphi^*_{\alpha}({\bf r})\varphi_{\beta}({\bf r})$, which is used in evaluating the xc potential from Eq.~(\ref{eq:lda}). The Hartree term in Eq.~(\ref{eq:ks}) is expressed through 
$\langle\alpha|v_{\rm H}({\bf r};[n_\sigma])|\beta\rangle=\sum_{i,\sigma}\sum_{\gamma,\delta}
[C^{i,\sigma}_\gamma]^*C^{i,\sigma}_\delta\,V_{\alpha\gamma\delta\beta}$, 
where
$
V_{\alpha\gamma\delta\beta}=
\int d^2{\bf r}\,d^2{\bf r}'\,v(|{\bf r}-{\bf r}'|)
\varphi^*_\alpha({\bf r})\,\varphi^*_\gamma({\bf r}')\,
\varphi_\delta({\bf r}')\,\varphi_\beta({\bf r})\,
$
are the two-body Coulomb matrix elements. Selection rules on the quantum numbers are hidden in $V_{\alpha\gamma\delta\beta}$: for instance, $V_{\alpha\gamma\delta\beta}$ vanishes unless the angular momentum is conserved in a scattering process, {\it i.e.} unless
$
M_{\alpha}+M_{\gamma}=M_{\delta}+M_{\beta}
$.
This is easily verified through an expansion of  $1/|{\bf r}-{\bf r}'|$ in cylindrical coordinates~\cite{cohl_1999}.

In practice, the sums over the FD basis elements must be truncated. 
The numerical calculations have used $N_{\rm max}=20$ energy levels, 
which corresponds to $(N_{\rm max}+1)(N_{\rm max}+2)/2=231$ single-particle states. 
Convergence of the self-consistent procedure has been achieved 
with a precision of at least $10^{-6}$ on the electron density.

\section{Testing the LSDA for the one-body densities}
\label{sect:ed}

A main aim of this section is to compare our LSDA results for the one-body density profiles
with the state-of-the-art DMC data of Pederiva {\it et al.}~\cite{pederiva_2000}. The comparison confirms the conclusions already drawn in Ref.~\onlinecite{pederiva_2000} and will give us confidence in the inputs to be used in our calculations of electron-pair correlations that will be reported in Sect.~\ref{sect:paircorrelation}. We also report numerical results obtained within the HF approximation~\cite{yannouleas_1999} and in some cases ($N=6$ and $30$) we illustrate the role of electron-electron interactions by showing the density profiles for noninteracting electrons ($\lambda=0$).

\begin{figure}
\begin{center}
\tabcolsep=0cm
\begin{tabular}{cc}
\includegraphics[width=0.50\linewidth]{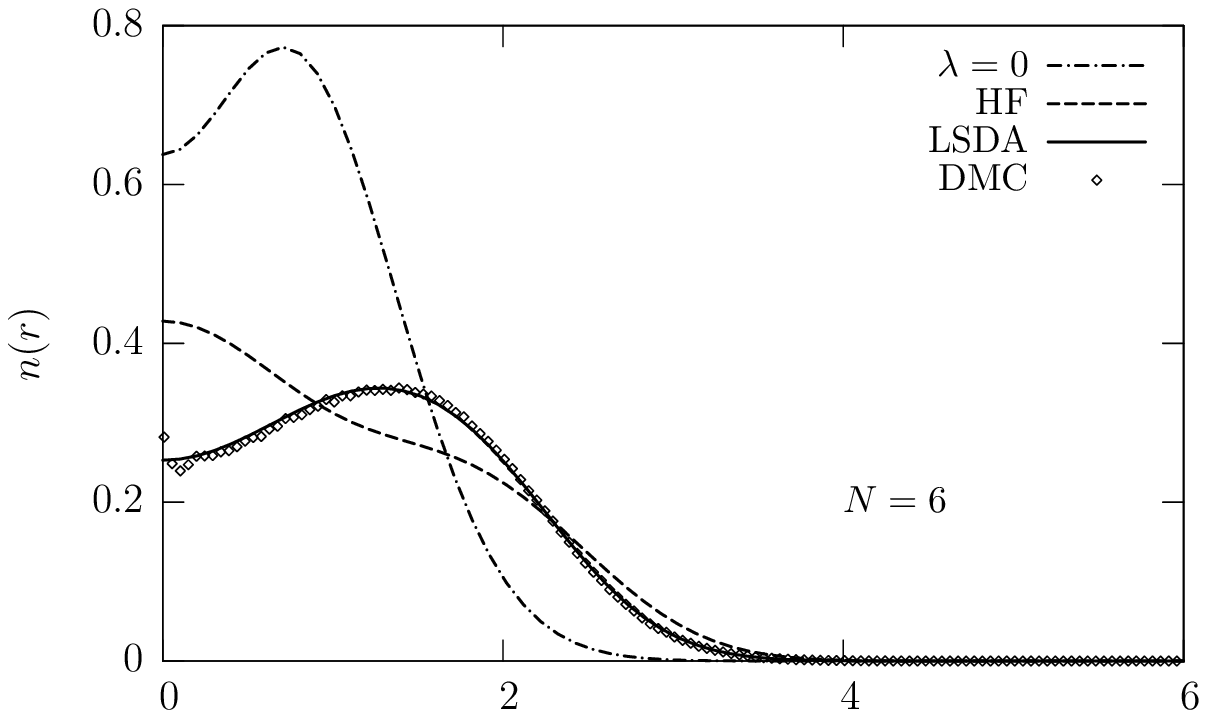}&
\includegraphics[width=0.50\linewidth]{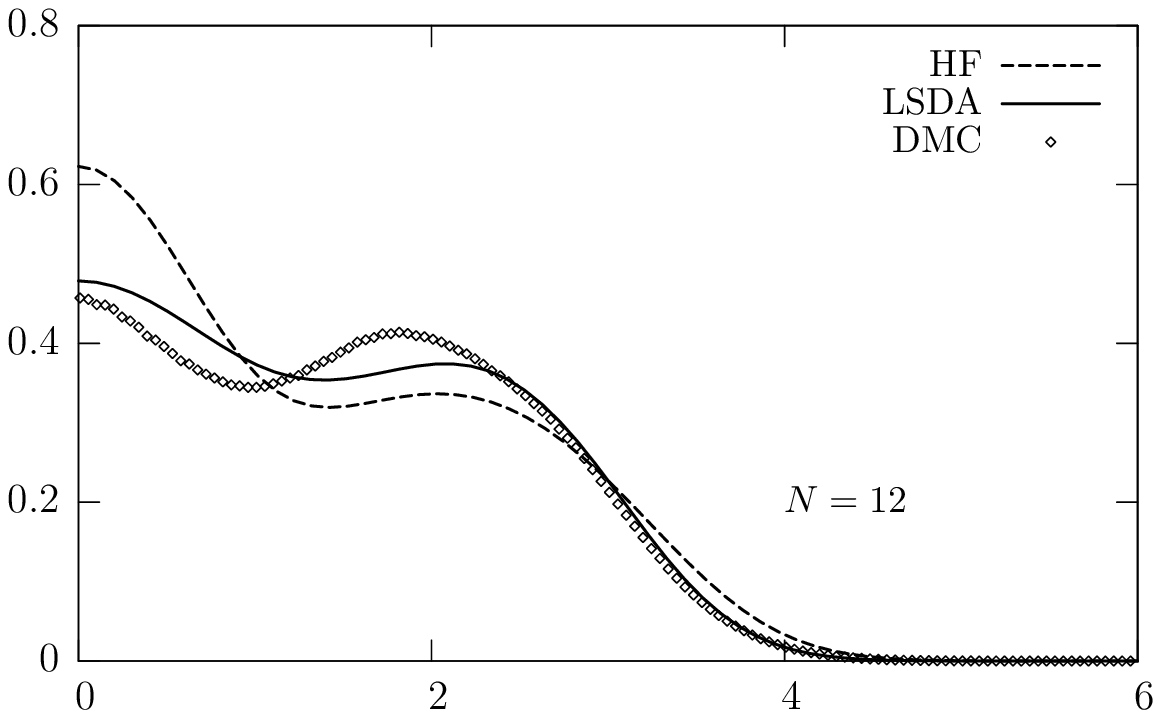}\\
\includegraphics[width=0.50\linewidth]{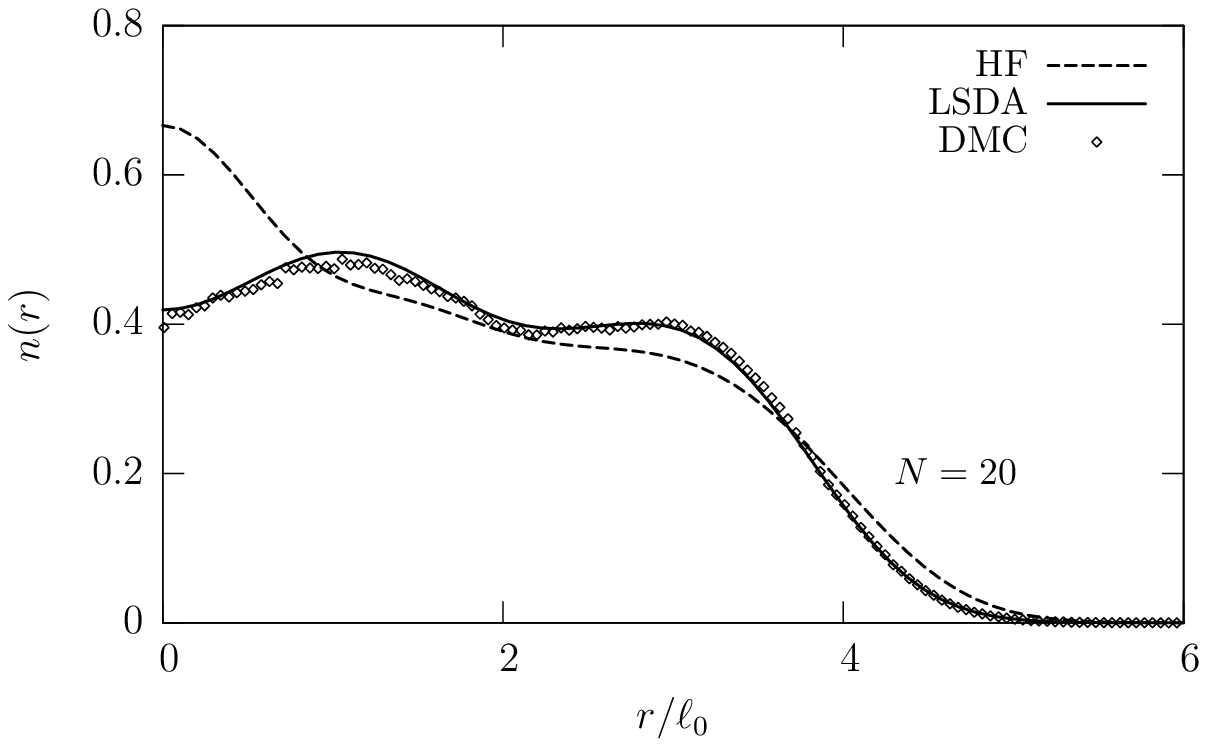}&
\includegraphics[width=0.50\linewidth]{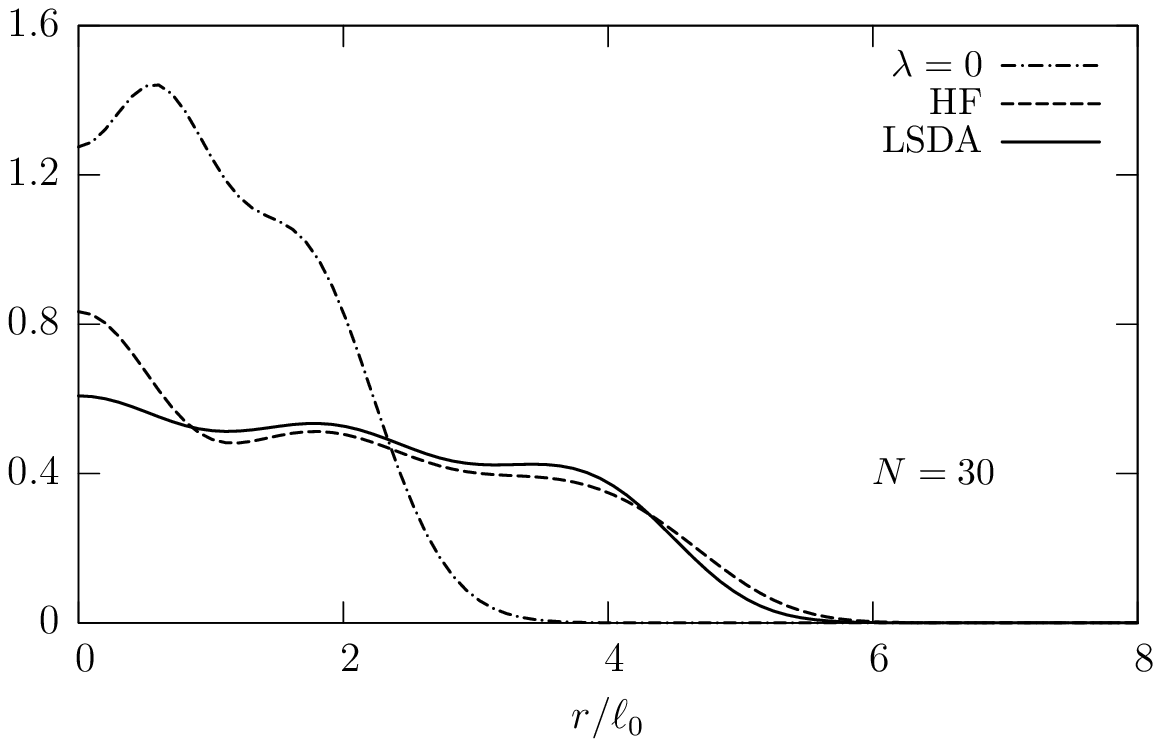}
\end{tabular}
\end{center}
\caption{Density profile $n(r)$ (in units of $\ell^{-2}_0$) as a function of $r/\ell_0$ 
for a paramagnetic QD with $N=6,12,20$ and $30$ electrons at $\lambda=1.89$. The results of the LSDA and of the HF 
are compared with the DMC data of Ref.~\onlinecite{pederiva_2000}. The dash-dotted lines are for noninteracting electrons.\label{fig:rho_para}}
\end{figure}

\begin{figure}
\begin{center}
\includegraphics[width=0.95\linewidth]{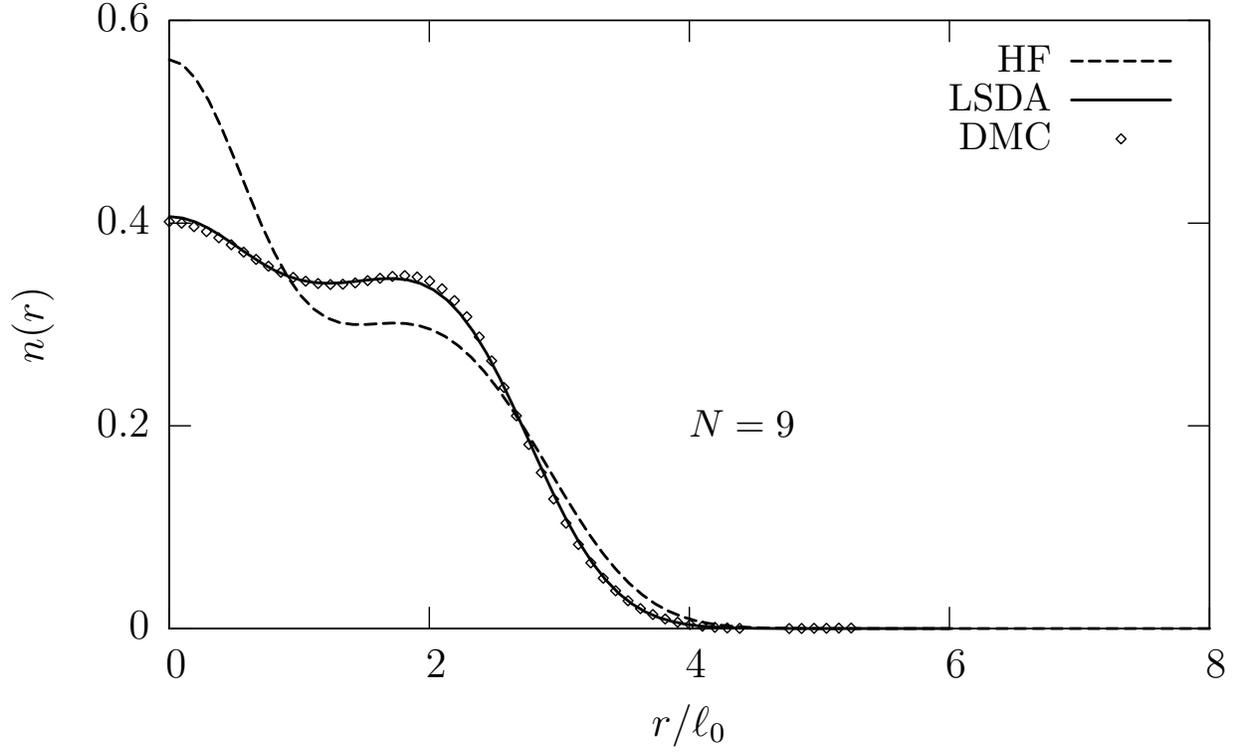}
\caption{Density profile $n(r)$ (in units of $\ell^{-2}_0$) as a function of $r/\ell_0$ for a partially spin-polarized QD with $N=9$ electrons at $\lambda=1.89$. The symbols are as in Fig.~\ref{fig:rho_para}.\label{fig:rho9b}}
\end{center}
\end{figure}

\begin{figure}
\begin{center}
\includegraphics[width=0.95\linewidth]{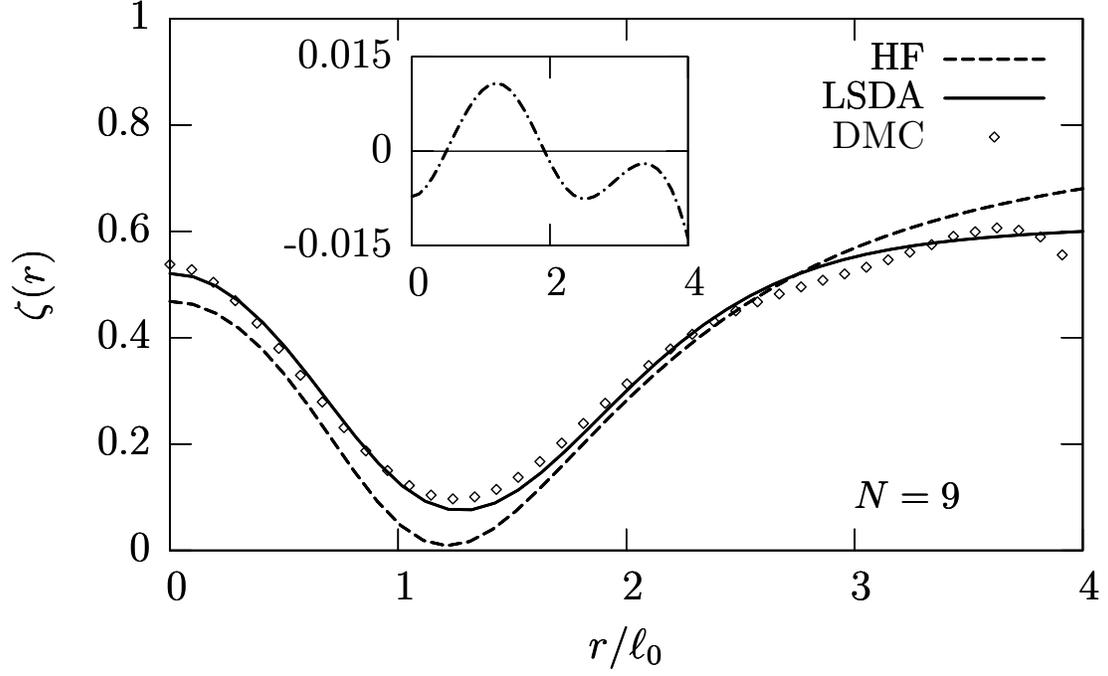}
\caption{Local spin polarization $\zeta(r)$ as a function of $r/\ell_0$ for 
a partially spin-polarized QD with $N=9$ electrons at $\lambda=1.89$. 
The symbols are as in Fig.~\ref{fig:rho_para}. 
The inset shows the difference between $\zeta(r)$ calculated with two different parametrizations for $\varepsilon^{\rm hom}_{\rm xc}(n,\zeta)$ (see text).\label{fig:rho9c}}
\end{center}
\end{figure}

Our LSDA and HF calculations of $n_\sigma(r)$ for circular $2D$ QD's refer to the cases $N=6,9,12,20$ and $30$. 
The confinement energy has been chosen as $\hbar \omega_0=3.32\,{\rm meV}$, which corresponds to $\lambda=1.89$. 
A summary of our main results is shown in Figs.~\ref{fig:rho_para}-\ref{fig:rho9c}. 

It is immediately evident from Fig.~\ref{fig:rho_para} that the LSDA density profiles are in excellent agreement with the DMC data, except for $N=12$. We have no explanation for this specific discrepancy. 
In the case $N=9$ the ground state at $\lambda=1.89$ is partially spin-polarized with $S_z=3/2$, while in all other cases it is paramagnetic ($S_z=0$). Figures~\ref{fig:rho9b} and~\ref{fig:rho9c} show that in the spin-polarized case at $N=9$ the agreement with the DMC data is excellent for both the total density profile $n(r)$ and the local spin polarization $\zeta(r)$. We have also checked that our LSDA results are not unduly sensitive to the input chosen for $\varepsilon^{\rm hom}_{\rm xc}(n,\zeta)$. We have tested in this respect the earlier parametrization of $\varepsilon^{\rm hom}_{\rm xc}(n,\zeta)$ given by Tanatar and Ceperley~\cite{tanatar_1989} and found minor differences arising in the local spin polarization for $N=9$, as is shown in the inset in Fig.~\ref{fig:rho9c}.  

We also confirm that the HF is not a good approximation for the ground-state density profiles, especially for small values of $N$ where the role of correlations is more important. The quality of the HF results appears to improve with increasing $N$, as indicated by the case $N=30$ in Fig.~\ref{fig:rho_para}. The local spin polarization for the case $N=9$ in Fig.~\ref{fig:rho9c} is also reasonably accounted for. A brief discussion of symmetry breaking in HF calculations of the one-body density is given in the Appendix.

\section{Pair correlations}
\label{sect:paircorrelation}

We have seen in Sect.~\ref{sect:sdft_acf} how the PDF enters the adiabatic connection formula 
for a formal definition of the xc energy functional. The PDF directly describes the conditional probability density 
$P({\bf r}',\sigma'|{\bf r},\sigma)=n_{\sigma'}({\bf r}')g_{\sigma\sigma'}({\bf r},{\bf r}')$ 
of finding an electron with spin $\sigma'$ at position ${\bf r}'$ when another electron with spin $\sigma$ 
is at position ${\bf r}$. In the homogeneous $2D$ electron fluid the increase of coupling strength with decreasing particle density towards a spin-polarized state and a triangular Wigner crystal is accompanied by strengthening short-range order in the electron-pair distribution. This is signaled by the emergence of a peak in $g(r)$ at a relative distance $r$ approaching the first-neighbor distance $d_{\rm WC}=(\sqrt{3}n/2)^{-1/2}$ in the crystal~\cite{tanatar_1989,attaccalite_2002}. 

Of course, the standard formulation of SDFT only gives access to the equilibrium one-body densities $n_{\sigma}({\bf r})$. Several attempts have been made~\cite{pdft} to build a generalized functional approach having 
the pair density $n^{(2)}_{\sigma\sigma'}({\bf r},{\bf r}')$ as its basic variable, from which both $n_{\sigma}({\bf r})$ and $g_{\sigma\sigma'}({\bf r},{\bf r}')$ may be obtained. A practicable self-consistent procedure to calculate $g_{\sigma\sigma'}({\bf r},{\bf r}')$ has been proposed by Davoudi {\it et al.}~\cite{davoudi_2002}, who extended to inhomogeneous fluids an approach originally used by Overhauser~\cite{overhauser_1995} to evaluate electron-pair correlations at contact. An Overhauser-type approach has also been set up~\cite{gori_giorgi&savin_2004} for calculating the angularly and center-of-mass averaged pair density, which suffices for evaluating the xc energy of an inhomogeneous electron system.

\subsection{The average spin-density approximation for the PDF}

In the present context, we examine an alternative approximate approach to the PDF, allowing relatively simple numerical calculations with results that will be compared with the DMC data of Pederiva {\it et al.}~\cite{pederiva_2000} for a $2D$ QD. Our approach is inspired to the so-called average-density and weighted-density approximations, that have been proposed in the literature for the purpose of transcending the LSDA in the evaluation of the xc energy functional (see Dreizler and Gross~\cite{dreizler_book} and references therein). These approximations satisfy by construction the ``central sum rule" in Eq.~(\ref{eq:central_sum_rule}).

In this so-called average-spin-density approximation (ASDA) the functional dependence of $g_{\sigma\sigma'}({\bf r},{\bf r}'; [n_\sigma])$ on $n_\sigma({\bf r})$ is taken in the form
\begin{equation}\label{eq:average}
g_{\sigma\sigma'}({\bf r},{\bf r}'; [n_\sigma])\simeq \left.g^{\rm \scriptstyle hom}_{\sigma\sigma'}(|{\bf r}-{\bf r}'|;n,\zeta)\right|_{n\rightarrow {\bar n}({\bf r},{\bf r}'),\zeta\rightarrow{\bar \zeta}({\bf r},{\bf r}')}
\equiv \left.g_{\sigma\sigma'}({\bf r},{\bf r}')\right|_{\rm ASDA}
\end{equation}
where
\begin{equation}
\left\{
\begin{array}{l}
{\bar n}({\bf r},{\bf r}')=[n({\bf r})+n({\bf r}')]/2\\
{\bar \zeta}({\bf r},{\bf r}')=[\zeta({\bf r})+\zeta({\bf r}')]/2
\end{array}
\right.
\end{equation}
(see also the work of Ebner {\it et al.}~\cite{ebner_1976} and of Yamashita and Ichimaru~\cite{yamashita_1984}). Contrary to the LSDA, the ASDA satisfies the symmetry property
$
g_{\sigma\sigma'}({\bf r},{\bf r}')=g_{\sigma'\sigma}({\bf r}',{\bf r})
$.
It still implies some minor violations of the central sum rule: 
for instance, we have verified that it may lead to deviations from the requirement in Eq.~(\ref{eq:central_sum_rule}) which become as large as $5\%$ for the spin-summed PDF in the bulk of a QD with $N=9$ and $\lambda=1.89$.

\begin{figure}
\begin{center}
\includegraphics[width=0.95\linewidth]{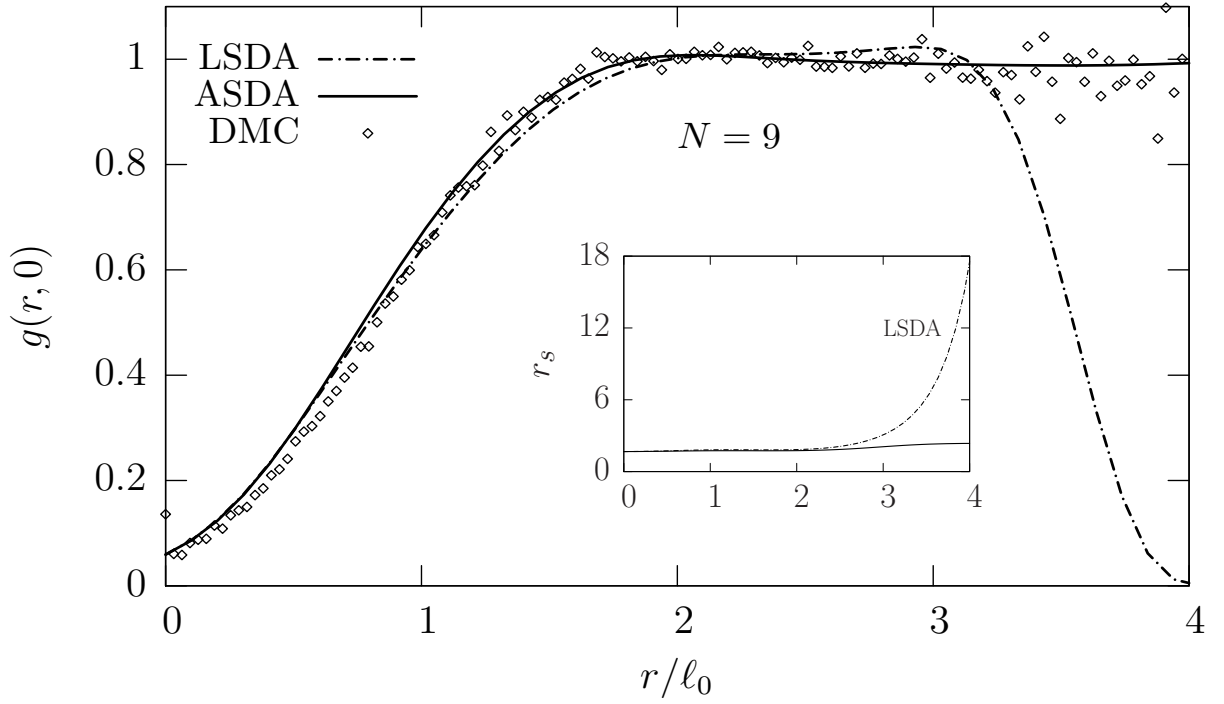}
\caption{Spin-summed PDF $g(r,0)$ as a function of $r/\ell_0$ for a 
partially spin-polarized QD with $N=9$ at $\lambda=1.89$. 
The results of the LSDA and ASDA are compared with the DMC data of Ref.~\onlinecite{pederiva_2000}. The inset shows $r_s^{\rm \scriptscriptstyle LSDA}(r)$ (dash-dotted line) and $r_s^{\rm \scriptscriptstyle ASDA}(r,0)$ (solid line) as functions of $r/\ell_0$ (see text).\label{fig:g9}}
\end{center}
\end{figure}

In Fig.~\ref{fig:g9} we compare our ASDA and LSDA results for the spin-summed PDF, defined as
\begin{equation}\label{eq:spin_summed}
g({\bf r},{\bf r}')=\sum_{\sigma,\sigma'}\frac{n_\sigma({\bf r})n_{\sigma'}({\bf r}')}{n({\bf r})n({\bf r}')}g_{\sigma\sigma'}({\bf r},{\bf r}')\,,
\end{equation}
with the DMC data of Pederiva {\it et al.}~\cite{pederiva_2000} on a QD with $N=9$ electrons and $\lambda=1.89$. 
The quantity being shown in Fig.~\ref{fig:g9} is $g({\bf r},{\bf r}'=0)$, which depends only on the modulus 
$r=|{\bf r}|$ owing to the circular symmetry of the ground-state density. As already noted, the ASDA satisfies the symmetry property of $g_{\sigma\sigma'}({\bf r},{\bf r}')$ and it does not matter whether one sets ${\bf r}$ or ${\bf r}'$ to zero (corresponding to the center of the QD). However this is not the case for the LSDA, and we have decided to show in Fig.~\ref{fig:g9} the choice that corresponds to setting ${\bf r}'=0$.

It is seen in Fig.~\ref{fig:g9} that the ASDA and the LSDA give essentially the same results when the second electron is also close to the center of the QD, and are in good agreement with the DMC data. But the LSDA badly fails in describing long-range correlations, because it breaks down across the edge of the QD at $r\simeq 3 \ell_0$ 
where $n(r)$ is rapidly dropping to zero. The ASDA is instead calculated at the average density 
${\bar n}(r,0)$, which tends smoothly to a constant across the QD edge. These behaviors can be emphasized by referring to local definitions of the $r_s$ density parameter as
$
r_s^{\rm \scriptscriptstyle LSDA}({\bf r})=[\pi n({\bf r})]^{-1/2}/a^\star_B
$
and 
$
r_s^{\rm \scriptscriptstyle ASDA}({\bf r}, {\bf r}')=[\pi {\bar n}({\bf r}, {\bf r}')]^{-1/2}/a^\star_B
$.
As is shown in the inset in Fig.~\ref{fig:g9}, while $r_s^{\rm \scriptscriptstyle ASDA}(r,0)$ remains essentially constant on crossing the QD edge, $r_s^{\rm \scriptscriptstyle LSDA}(r)$ increases in an exponential way heralding the breakdown of the LSDA. In practice, however, this breakdown is less serious than it may seem, since the two-body density is determined by the PDF multiplied by density factors.

We proceed to present a broader view of the ASDA spin-averaged PDF for the same partially spin-polarized QD. 
Figure~\ref{fig:inset} shows the geometrical coordinates that will be used in the following figures.
In Fig.~\ref{fig:g9_3D} we show a three-dimensional plot of $g({\bf r},{\bf r}')$, when both ${\bf r}=x\,{\hat {\bf x}}$ and ${\bf r}'=x'\,{\hat {\bf x}}$ lie on a line ${\hat {\bf x}}$ passing through the center of the confining potential (see Fig.~\ref{fig:inset}, left). The main features in Fig.~\ref{fig:g9_3D} are as follows: 
(i) 
the Pauli-Coulomb hole lying along the diagonal $x=x'$; 
(ii) 
the correlation-induced oscillations which are seen to lie along directions parallel to this diagonal, as are better seen in the contour plot; 
and 
(iii) 
the essentially flat asymptotic regions further out. 
We may remark that $r_s^{\rm \scriptscriptstyle ASDA}(x,x')$ is a smooth and bounded function if at least one of the two coordinates lies in the bulk of the QD inside its edge. The calculation loses meaning when both coordinates are far outside the edge, so that ${\bar n}(x,x')$ is rapidly vanishing. In fact, the needed input on the PDF of the homogeneous $2D$  EG from Ref.~\onlinecite{gori_giorgi_2004} is limited to electron densities corresponding to $r_s=(\pi n)^{-1/2}/a^\star_B$ up to the value $40$.
The contour plot in Fig.~\ref{fig:g9_3D} shows as hatched areas these regions of inapplicability, located approximately at $(|x|, |x'|)>3 \ell_0$.

\begin{figure}
\begin{center}
\includegraphics[width=0.80\linewidth]{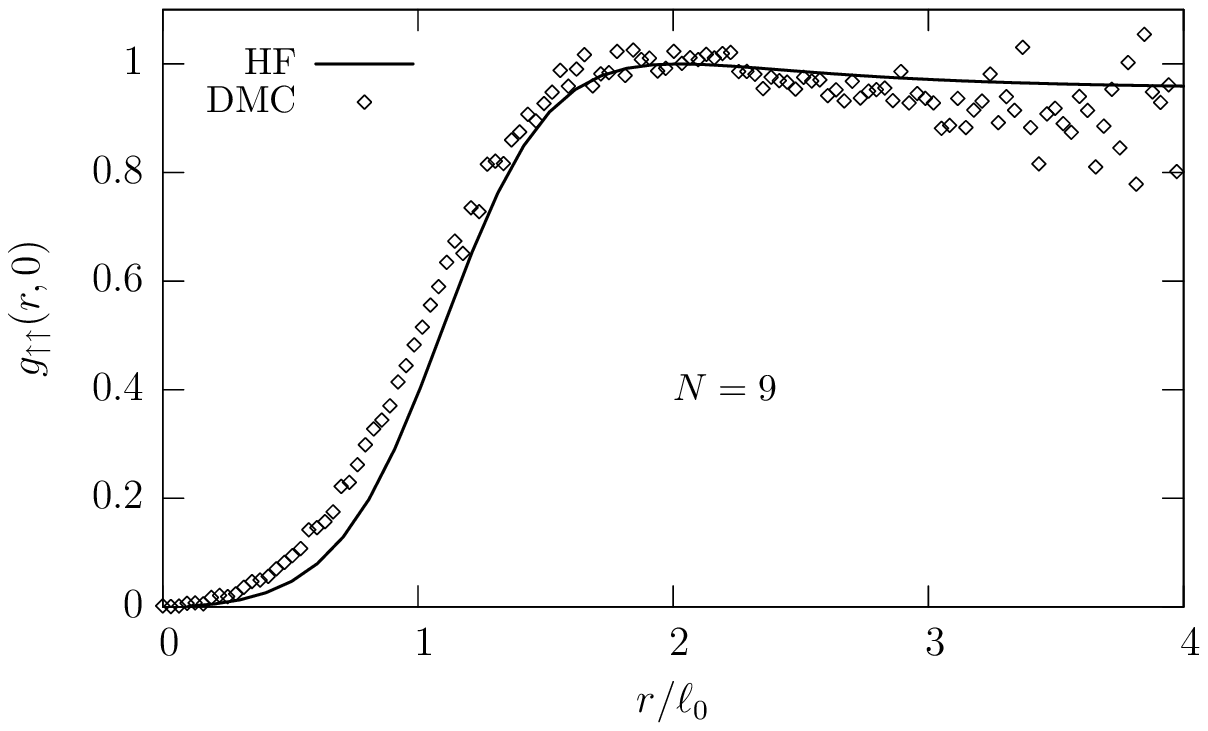}\\
\includegraphics[width=0.80\linewidth]{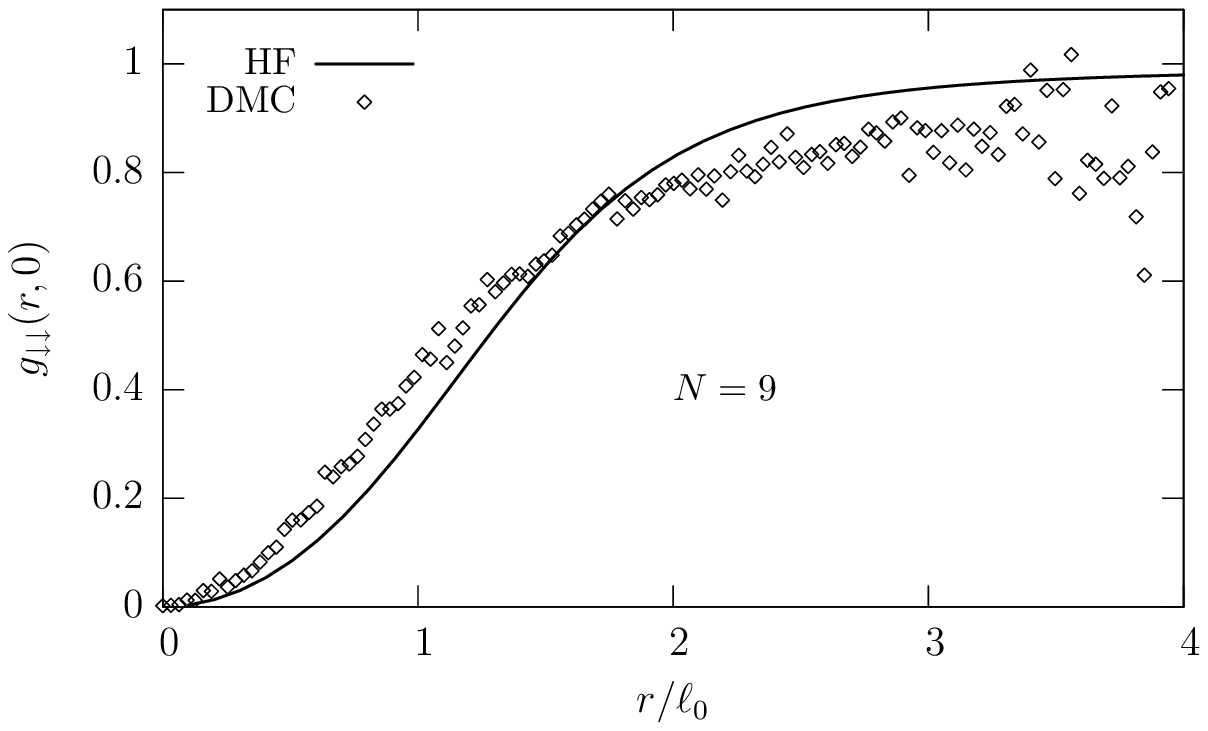}
\caption{Parallel-spin PDF $g_{\sigma\sigma}(r,0)$ as a function of $r/\ell_0$ for a partially 
spin-polarized QD with $N=9$ electrons at $\lambda=1.89$. The results of the Hartree-Fock approximation are compared with the DMC data of Ref.~\onlinecite{pederiva_2000}.\label{fig:hf_spin_resolved}}
\end{center}
\end{figure}

Before concluding this section we should comment on the spin-resolved pair correlations for the same QD. 
Unfortunately Ref.~\onlinecite{gori_giorgi_2004} does not provide analytical representations for the spin-resolved PDF of the homogeneous $2D$ EG at finite values of $\zeta$. This has prevented us from building the correspondent spin-resolved ASDA PDF for partially spin-polarized QD's.
However, in Fig.~\ref{fig:hf_spin_resolved} we compare the parallel-spin PDF's in the HF approximation, defined as
\begin{equation}
\left.g_{\sigma\sigma}({\bf r},{\bf r}')\right|_{\rm HF}=1-\frac{\rho_{\sigma\sigma}({\bf r},{\bf r}')\rho_{\sigma\sigma}({\bf r}',{\bf r})}{n_{\sigma}({\bf r})n_{\sigma}({\bf r}')}\,,
\end{equation}
with the DMC data of Pederiva {\it et al.}~\cite{pederiva_2000}. The quantity being shown in this figure is $\left.g_{\sigma\sigma}(r,0)\right|_{\rm HF}$. We conclude that at the value of the coupling strength in Fig.~\ref{fig:hf_spin_resolved} the parallel-spin HF PDF is already in fairly good agreement with the DMC results. On the other hand, the HF approximation completely misses antiparallel-spin electron-pair correlations by giving $\left.g_{\uparrow\downarrow}({\bf r},{\bf r}')\right|_{\rm HF}=1$.

\subsection{Evolution of short-range order towards Wigner-like order in a $2D$ QD}
\label{subsect:crystal}

The attainment of Wigner-like order in confined electronic system has been studied by a number of authors. In particular, Egger {\it et al.}~\cite{egger_1999} have reported a crossover from Fermi liquid to ``Wigner molecule" structure in a finite-temperature study of QD's containing up to $N=8$ electrons by path-integral Monte Carlo simulation (PIMC). A later PIMC study by Filinov {\it et al.}~\cite{filinov_2001} has regarded electron clusters in QD's with different particle numbers at various temperatures and coupling strengths. For even values of $N$ these authors took the electronic system in a paramagnetic state and predicted a ``phase diagram" at zero temperature, which shows a transition to a radially ordered state followed at much higher coupling strength by a transition to an angularly ordered state. For instance, in the case $N=10$ (that is the lowest value of the particle number in their study) the first transition occurs at $\lambda\simeq 20$ and the second at $\lambda \simeq 2770$. The good predictive value of the ASDA for electron-pair correlations allows us to inspect how the short-range order in a QD at weak coupling evolves with increasing $\lambda$ towards radial Wigner-like ordering. We do this below for the case $N=6$ and the results are presented in Figs.~\ref{fig:prob}-\ref{fig:angular}.

In calculating the one-body radial probability density $2\pi rn(r)$ we enforce circular symmetry and consider only spin states that are paramagnetic ($S_z=0$) or ferromagnetic ($S_z=3$). The ground state changes spin polarization with increasing coupling strength: for $\lambda=1.89$ and $3.54$ the paramagnetic state is lower in energy, but lies above the ferromagnetic state for $\lambda=6.35,10$, and $12$. Figure~\ref{fig:prob} shows the 
probability density for both states of spin polarization at the above values of the coupling strength. A shoulder and ultimately a marked minimum appear in the ferromagnetic state with increasing $\lambda$: similar results have already been reported by Egger {\it et al.}~\cite{egger_1999} and by Reimann {\it et al.}~\cite{reimann_2000}. The electronic system acquires the so-called $(1,5)$ structure consisting of one electron at the center of the trap and a surrounding ring of five electrons. We have checked that $\int_0^{r_{\rm min}}2\pi r\,n(r)dr =1$ for $\lambda=10$ and $12$, where $r_{\rm min}$ is the position of the minimum in the probability density, and found that the height $\Delta(\lambda)$ of the probability density at $r_{\rm min}$ vanishes for $\lambda\simeq 14$ (see the inset in Fig.~\ref{fig:prob}).

We turn to present the ASDA results for the evolution of the radial and angular dependence of the spin-summed PDF with increasing $\lambda$. Figure~\ref{fig:g_lambda} reports the function $g(r,0)$ for the ground state of the QD and shows that, whereas the paramagnetic ground state at weak coupling does not possess any pronounced radial structure, the ferromagnetic ground state at $\lambda=10$ and $12$ exhibits a main first-neighbor peak followed by secondary structures. All these structures are in phase with structures in the local coupling strength $r_s^{\rm \scriptscriptstyle ASDA}(r,0)$, as is shown in the inset in Fig.~\ref{fig:g_lambda}.

The growth of radial ordering with increasing $\lambda$ in the ferromagnetic ground state is even more clearly displayed by plotting the total conditioned probability density $2\pi r\, P(r|r'=0)\equiv 2\pi r\,n(r)g(r,0)$, which carries information on both the one-body density distribution and the radial electron-pair correlations. This function is shown in Fig.~\ref{fig:cond_proba} for the QD under discussion. The value of the coupling strength $\lambda\simeq 14$ at which the height of the minimum in this function vanishes represents within the present theory our estimate for the location of the transition to radial Wigner-like ordering in a parabolic QD containing six spin-polarized electrons.

Finally, the angular dependence of the electron-pair correlations in the ground state is illustrated in Fig.~\ref{fig:angular} at the radial distance $r_{\rm max}$ corresponding to the location of the absolute maximum in the probability density. The function that is being plotted at various values of $\lambda$ is 
$
g(\theta)=g^{\rm hom}(2r_{\rm max}\sin{(\theta/2)};n,\zeta)
$ 
evaluated at $n=n(r_{\rm max})$ and $\zeta=\zeta(r_{\rm max})$, with $\theta$ being the angle between ${\bf r}$ and ${\bf r}'$ as shown in the right panel of Fig.~\ref{fig:inset}. Of course, the mapping between $g(\theta)$ and the PDF of the homogeneous $2D$ EG is a consequence of the ASDA. Even at $\lambda=12$ the angular ordering of the electronic system in the QD is seen from Fig.~\ref{fig:angular} to be still very much liquid-like. Starting in the ferromagnetic state from an electron at $\theta=0$ on a circle at $r=r_{\rm max}$, we find an enhanced probability of having a first neighbor on each side of it and two additional structures further out on the circle, but there is no evidence for an ordered fivefold ring of electrons. Indeed, the positions of the peak structures in $g(\theta)$ are far from corresponding to regular pentagonal angles as would be appropriate for an angularly ordered $(1,5)$ structure.

As a final remark we notice that, while the angular distance $\theta_p(\lambda)$ from the first-neighbor peaks in 
Fig.~\ref{fig:angular} decreases with increasing $\lambda$, the preferred first-neighbor distance $d(\lambda)=2r_{\rm max}(\lambda)\sin{(\theta_p(\lambda)/2)}$ along the circle at $r_{\rm max}$ is increasing with $\lambda$. This is shown in the inset in Fig.~\ref{fig:angular} and is due to the increase in $r_{\rm max}$ with increasing Coulomb repulsions.

\section{Summary and conclusions}
\label{sect:conclusions}

In summary, the main original parts of this work have concerned the theory of the short-range order that may be met in electron assemblies confined inside $2D$ parabolic quantum dots in a weak-to-intermediate range of coupling strength. We have proposed a very practical scheme for the calculation of the pair distribution functions in these inhomogeneous electron systems and examined in great detail its predictions in two specific cases. For a partially spin-polarized system of nine electrons we have seen that the theory is able to quantitatively account for the available Diffusion Monte Carlo data on spin-averaged two-body correlations. We have added to this a panoramic view of the charge-charge correlations, that waits to be tested in further Monte Carlo studies.

The second problem that we have examined in detail has been the state of spatial short-range order in the paramagnetic and ferromagnetic states of a system of six electrons as a function of the coupling strength parameter. Naturally enough, by being based on a density functional approach that takes input from the homogeneous $2D$ electron gas, our predictions parallel to some extent the well-known phase behavior of this macroscopic system. On increasing the coupling strength the ground-state of the quantum dot first changes from paramagnetic to ferromagnetic and then acquires radial order in coexistence with orientational liquid-like short-range order, over the range of coupling strength that we have considered. It would be important, we feel, to re-examine these correlation properties in the quantum dot with six electrons by exact-diagonalization methods. One could test in this way to what extent the predictions that take their start from the macroscopic electron gas are in accord with those that are based on a few-electrons atomic viewpoint.

\acknowledgments
This work was partially supported by MIUR through PRIN2003. 
We acknowledge useful discussions with Prof. M. Bonitz, Dr. P. Gori-Giorgi, Prof. A. Kievsky, and Dr. S. Simonucci. 
We are grateful to Dr. F. Pederiva for providing us with the results of the Diffusion Monte Carlo studies.

\appendix*  

\section{Some comments on broken symmetry states}

An approximate treatment of a strongly correlated many-body problem can in principle lead to states with spontaneously broken rotational symmetry (see for instance the review of Reimann and 
Manninen~\cite{addition_energy} and the discussion given by Ring and Schuck~\cite{ring_1980}). A well studied example in the area of QD's is the self-consistent spin-and-space unrestricted HF treatment of the one-body density, which has been shown~\cite{yannouleas_1999} to break the rotational symmetry at relatively low values of the coupling strength. 

\begin{figure}
\begin{center}
\includegraphics[width=0.95\linewidth]{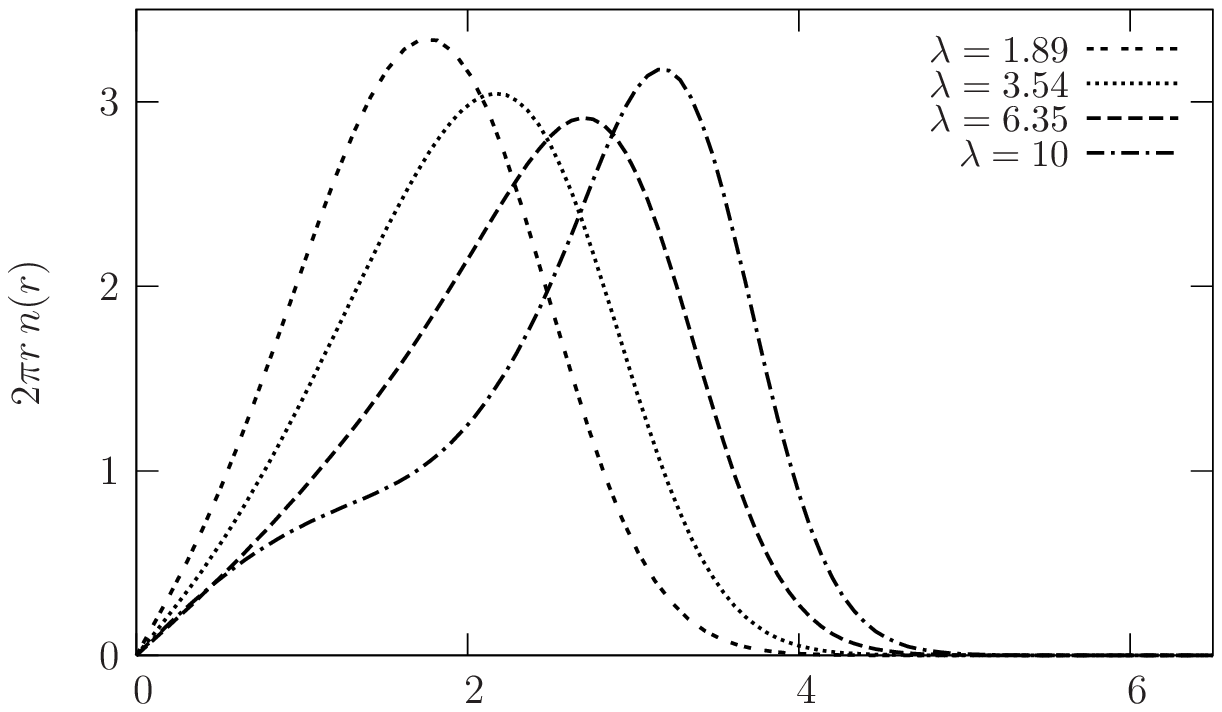}
\includegraphics[width=0.95\linewidth]{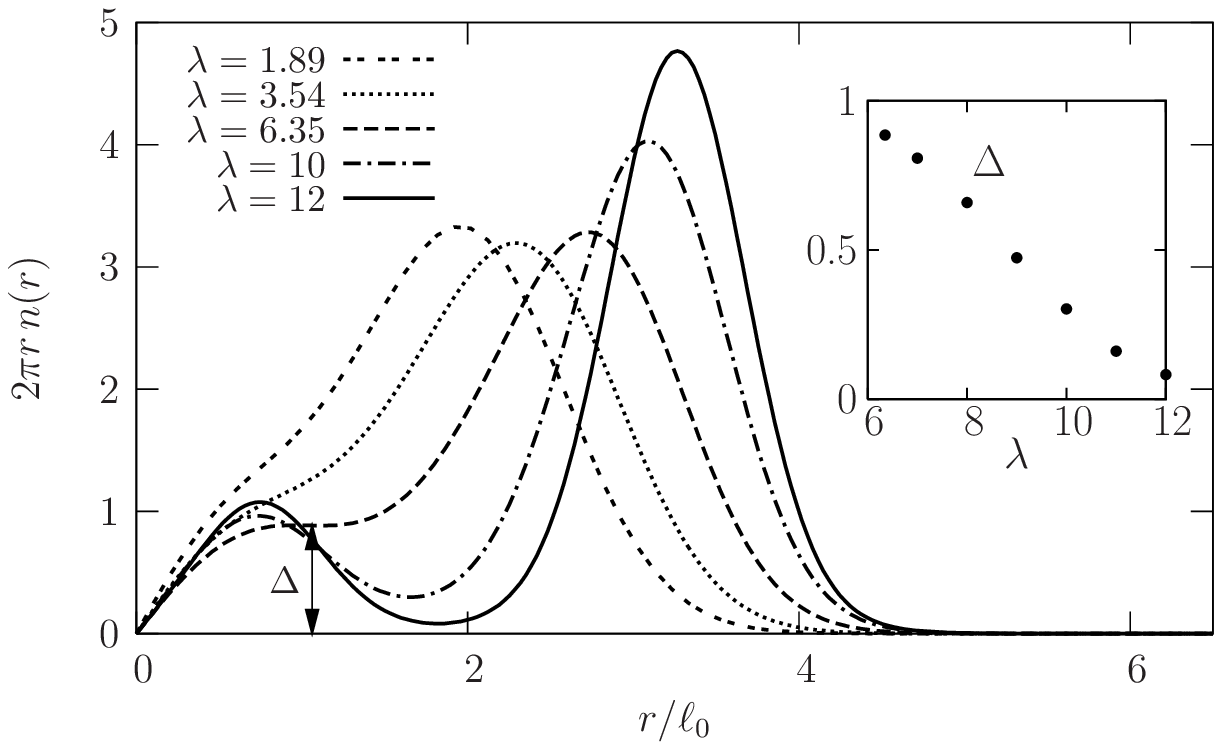}
\caption{Probability density (in units of $\ell^{-1}_0$) as a function of $r/\ell_0$ for a QD with $N=6$ electrons at varying $\lambda$: profiles for the paramagnetic state (top) and for the ferromagnetic state (bottom). The inset shows the height $\Delta$ of the minimum (in units of $\ell^{-1}_0$) as a function of $\lambda$.\label{fig:prob}}
\end{center}
\end{figure}

\begin{figure}
\begin{center}
\includegraphics[width=0.95\linewidth]{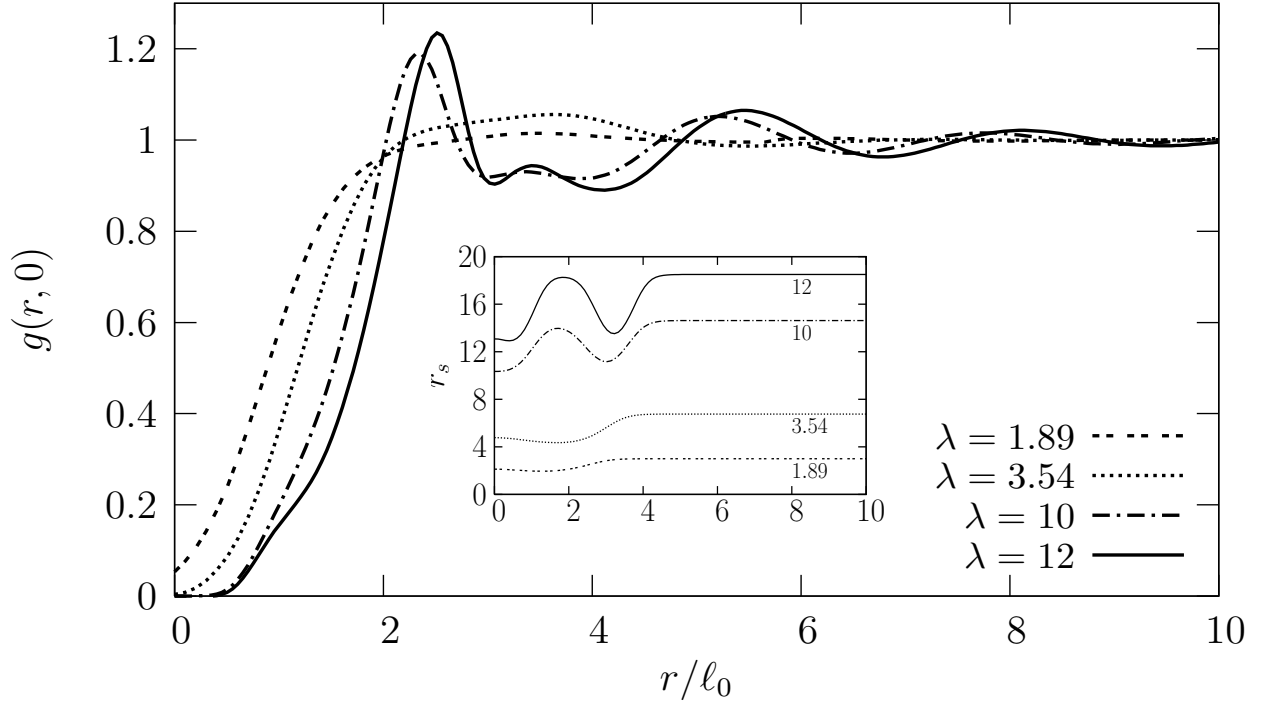}
\caption{Spin-summed PDF $g(r,0)$ as a function of $r/\ell_0$ for a QD with $N=6$ electrons in its ground state at various $\lambda$. The curves for $\lambda=1.89$ and $3.54$ refer to the paramagnetic state, 
while those for $\lambda=10$ and $12$ refer to the ferromagnetic state. The inset shows $r_s^{\rm \scriptscriptstyle ASDA}(r,0)$.\label{fig:g_lambda}}
\end{center}
\end{figure}

\begin{figure}
\begin{center}
\includegraphics[width=0.90\linewidth]{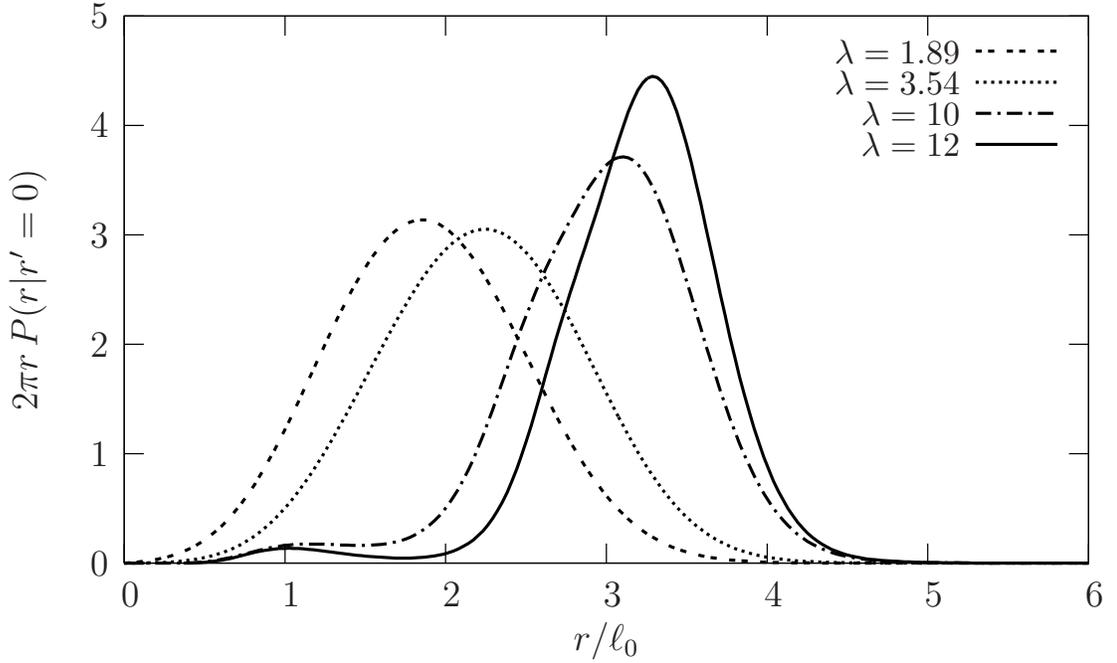}
\caption{Total conditioned probability density $2\pi r\, P(r|r'=0)$ as a function of $r/\ell_0$ for the ground state 
of a QD with $N=6$ electrons. The curves for $\lambda=1.89$ and $3.54$ refer to the paramagnetic state, 
while those for $\lambda=10$ and $12$ refer to the ferromagnetic state.\label{fig:cond_proba}}
\end{center}
\end{figure}

\begin{figure}
\begin{center}
\includegraphics[width=0.95\linewidth]{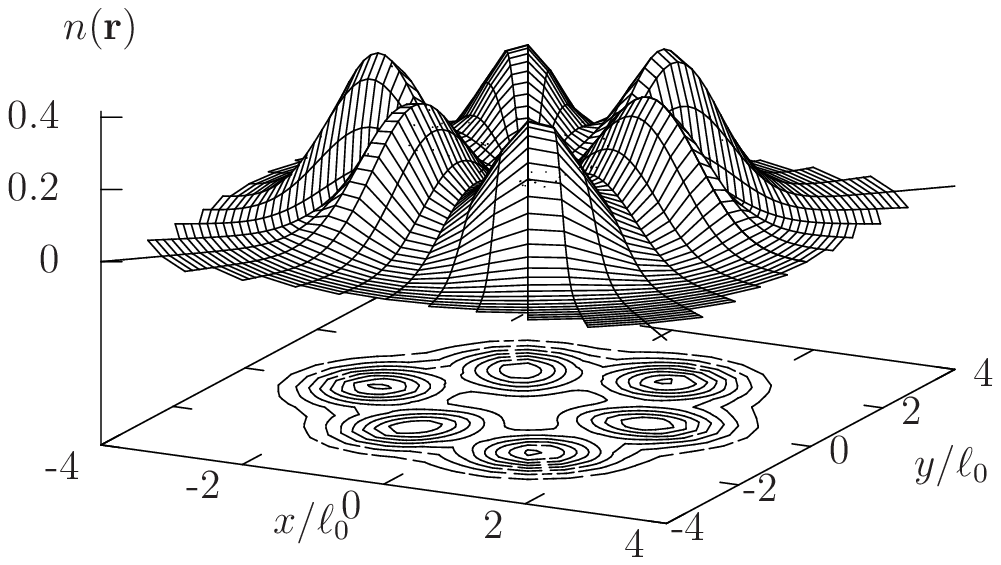}
\includegraphics[width=0.95\linewidth]{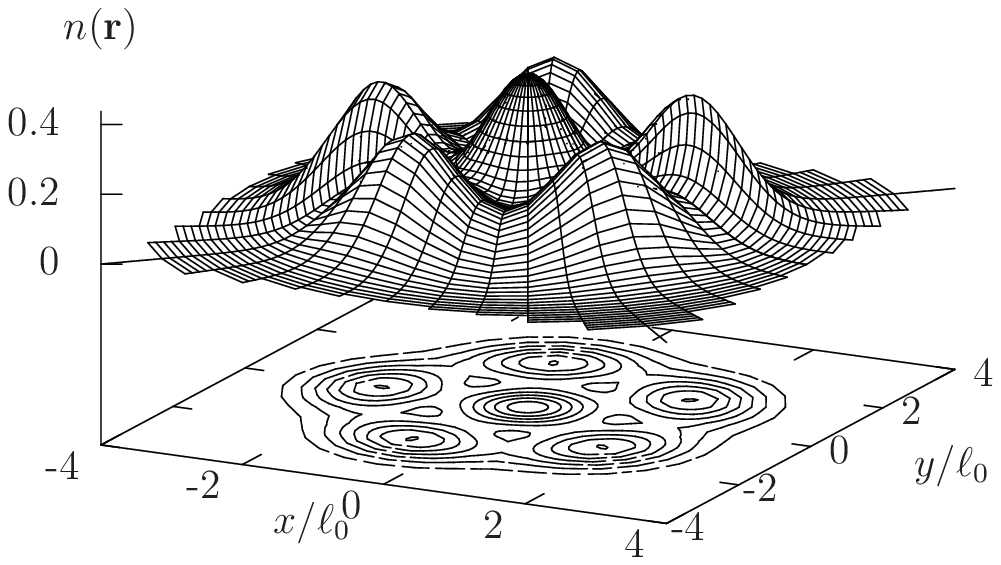}
\caption{Rotationally broken HF one-body density (in units of $\ell^{-2}_0$) for a QD with $N=6$ electrons at $\lambda=3.18$, in the paramagnetic (upper panel) and ferromagnetic (lower panel) case. At this value of 
$\lambda$ the HF approximation predicts the ferromagnetic state to lie at lower energy.\label{fig:hf}}
\end{center}
\end{figure}

As an illustrative example we show in Fig.~\ref{fig:hf} the HF one-body density for a QD with $N=6$ electrons at $\lambda=3.18$ in both the paramagnetic and the ferromagnetic case, in full agreement with the findings of Ref.~\onlinecite{yannouleas_1999}. The state of order that these pictures suggest for the six-electron system is very different from the results that we have found from our calculations of the electron-pair correlations, as reported in Figs.~\ref{fig:prob}-\ref{fig:angular}.

\newpage

\begin{figure}
\begin{center}
\tabcolsep=0 cm
\begin{tabular}{cc}
\includegraphics[height=5cm]{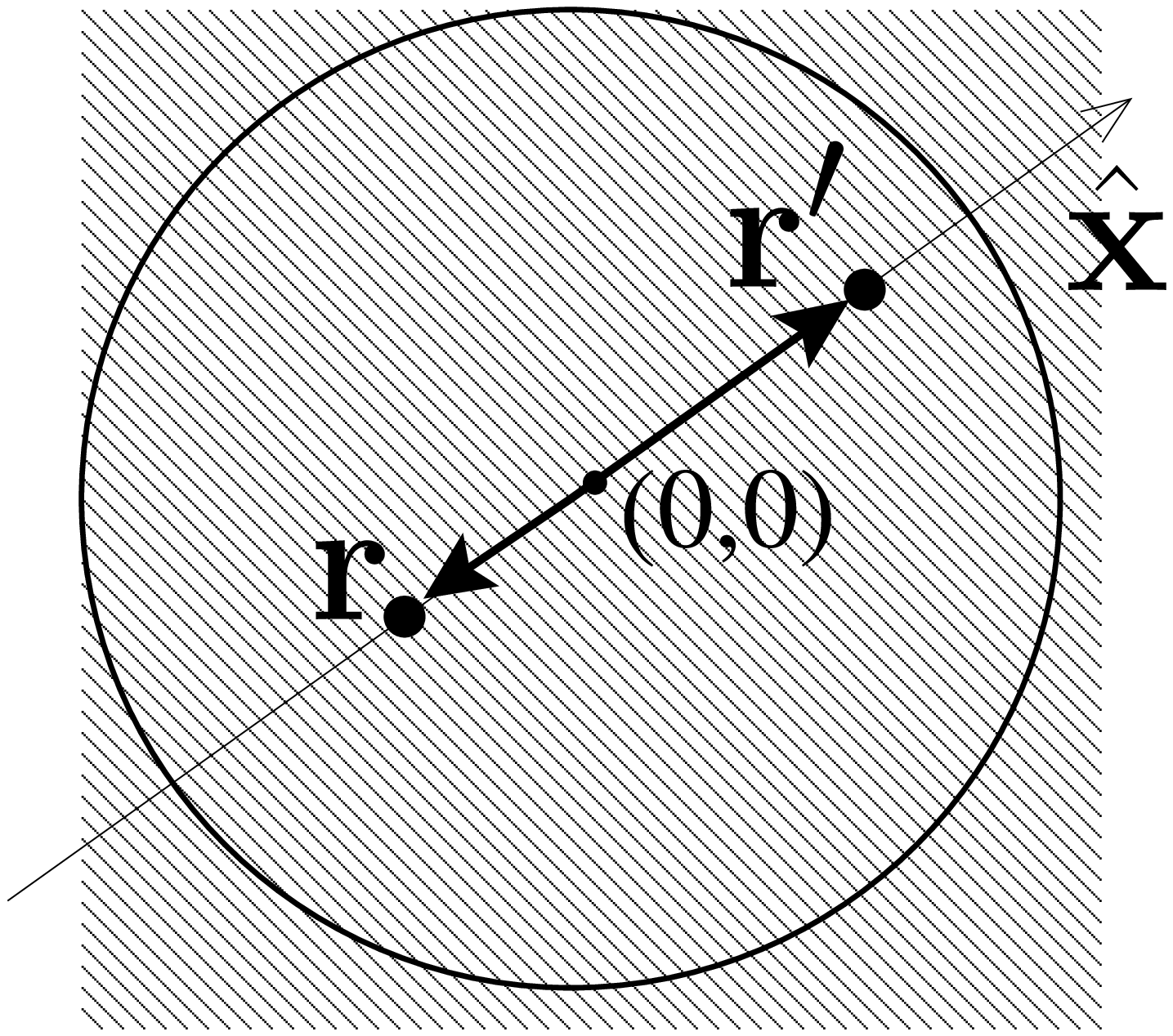}&
\includegraphics[height=5cm]{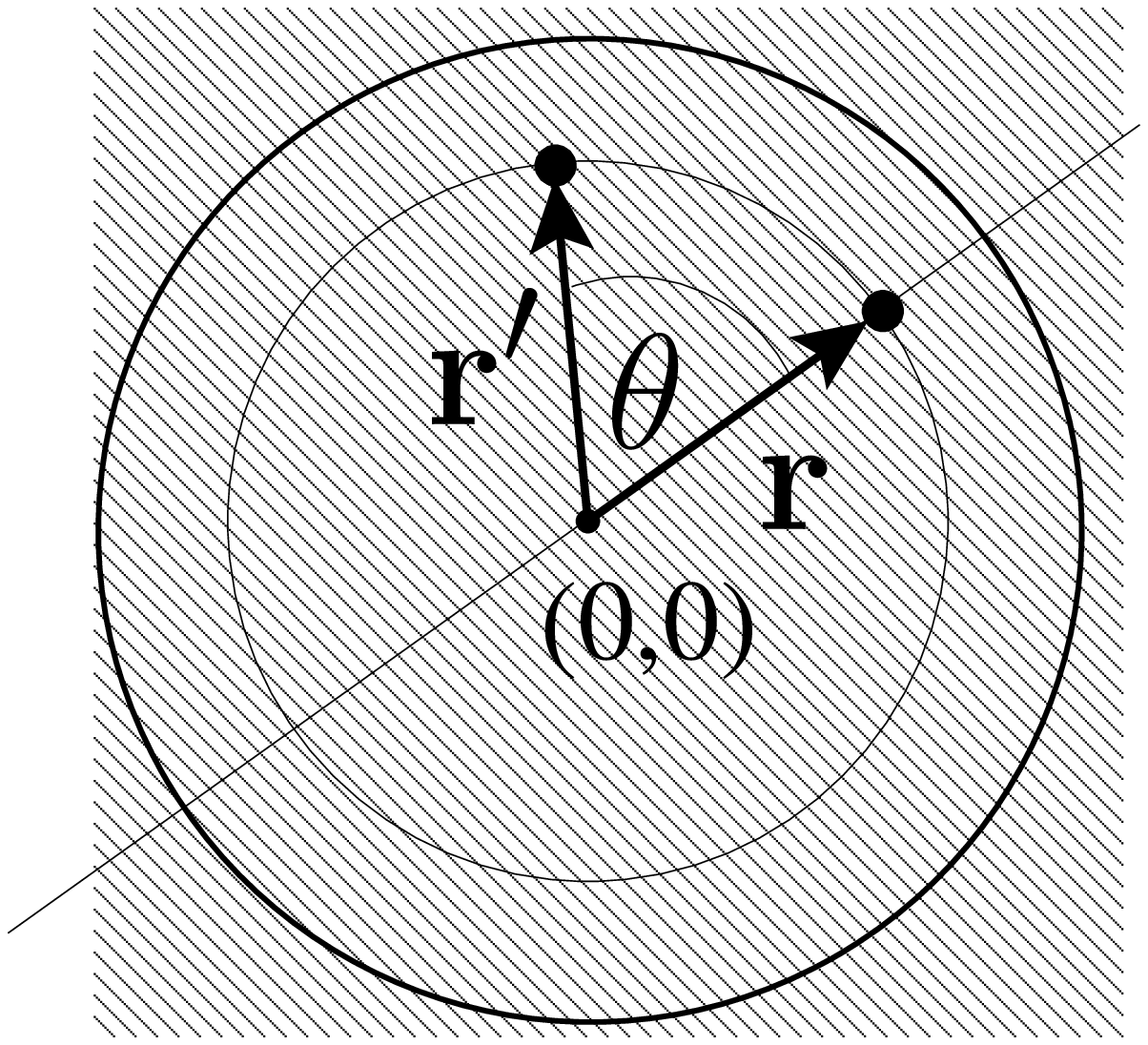}
\end{tabular}
\caption{Left: definition of $x$ and $x'$ appearing on the horizontal axis of Fig.~\ref{fig:g9_3D} (${\bf r}=x{\hat {\bf x}}$ and ${\bf r}'=x'{\hat {\bf x}}$). Right: definition of the azimuthal angle $\theta$ giving the abscissa 
in Fig.~\ref{fig:angular}.\label{fig:inset}}
\end{center}
\end{figure}

\begin{figure}
\begin{center}
\includegraphics[width=0.80\linewidth]{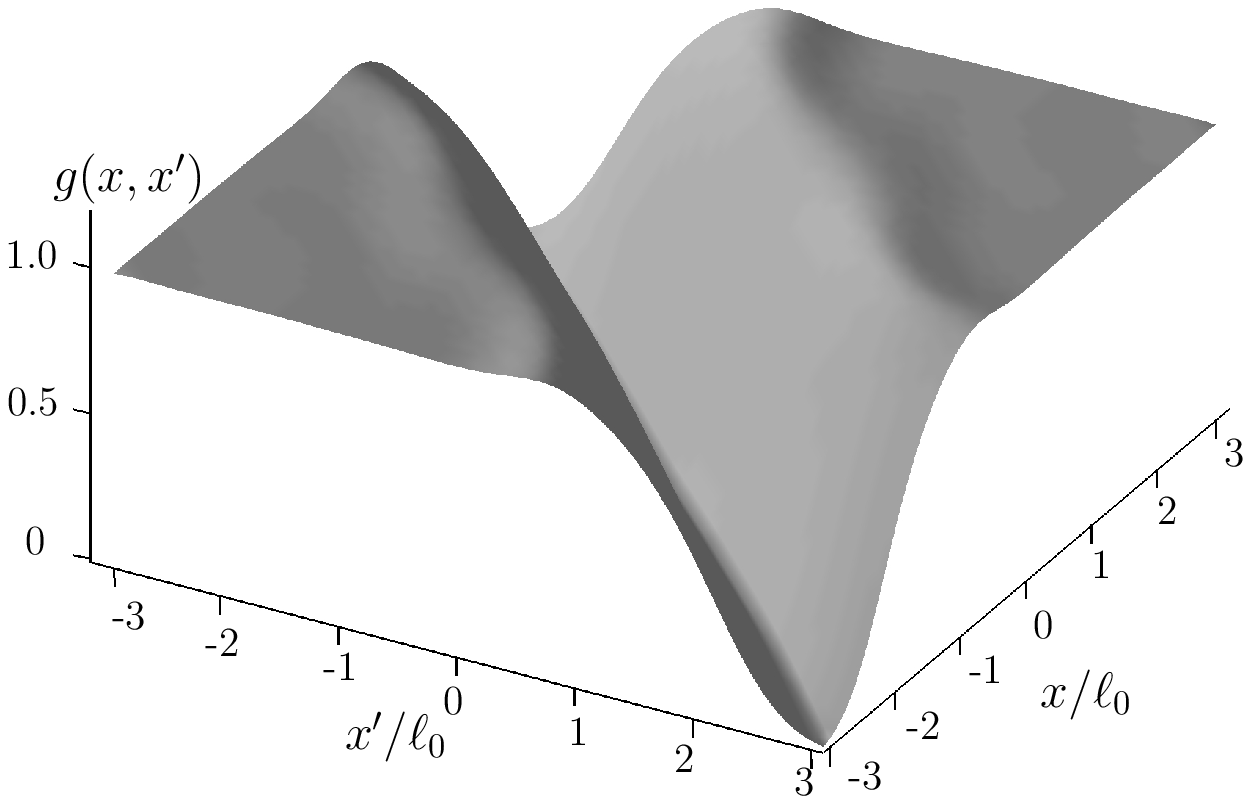}\\
\includegraphics[width=0.80\linewidth]{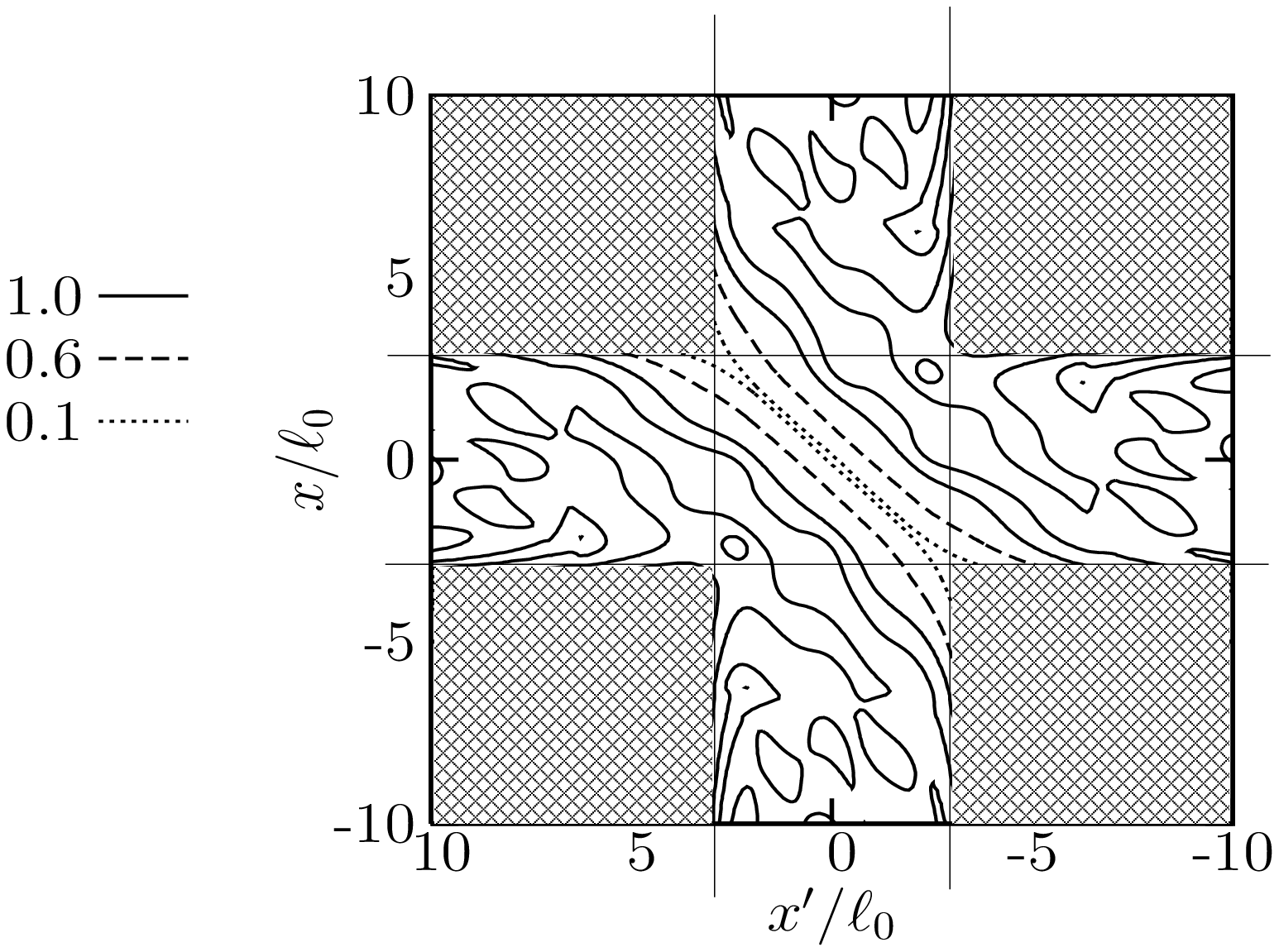}
\caption{Spin-summed PDF $g(x,x')$ as a function of $x/\ell_0$ and $x'/\ell_0$ for a partially 
spin-polarized QD with $N=9$ electrons at $\lambda=1.89$. The 
bottom panel shows a contour plot of $g(x,x')$: the thin solid lines limit the bulk central region 
$(|x|, |x'|)\lesssim 3\ell_0$ which is shown in the top panel.\label{fig:g9_3D}}
\end{center}
\end{figure}

\begin{figure}
\begin{center}
\includegraphics[width=0.90\linewidth]{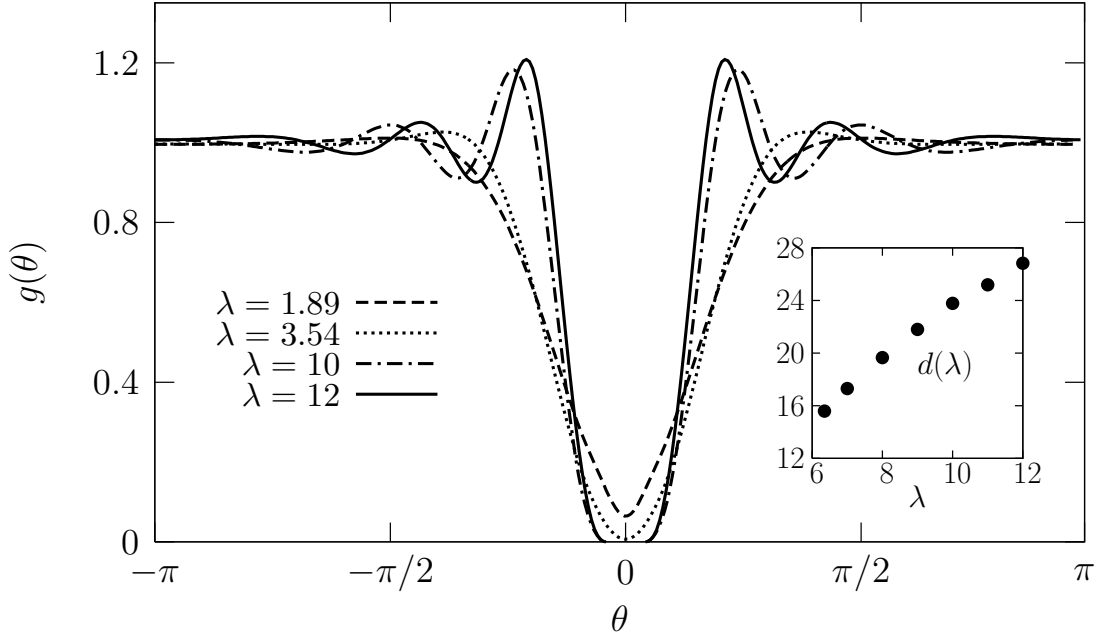}
\caption{Azimuthal plot of the spin-summed PDF as a function of the angle $\theta$ defined in Fig.~\ref{fig:inset}, 
for the ground state of a QD with $N=6$ electrons. The reference electron is located at $\theta=0$. 
The curves for $\lambda=1.89$ and $3.54$ refer to the paramagnetic state, while those for $\lambda=10$ and $12$ 
refer to the ferromagnetic state. The inset shows the preferred first-neighbor distance $d(\lambda)$ along the circle at $r=r_{\rm max}$, in units of $a^\star_B$.\label{fig:angular}}
\end{center}
\end{figure}

\begin{thebibliography}{99}
\bibitem{bec}
	See for instance C.J. Pethick and H. Smith, {\it Bose-Einstein Condensation in Dilute Gases} 
	(Cambridge University Press, Cambridge, England, 2002);
	A. Minguzzi, S. Succi, F. Toschi, M.P. Tosi, and P. Vignolo, Phys. Rep. {\bf 395}, 223 (2004).
\bibitem{deHeer}
	See for instance W. de Heer, \rmp {\bf 65}, 611 (1993); M. Brack, {\it ibid.} {\bf 65}, 677 (1993).
\bibitem{general_qd}
	L. Jacak, P. Hawrylak, and A. W{\'o}js, {\it Quantum Dots} (Springer, Berlin, 1998); 
	T. Chakraborty, {\it Quantum Dots: A Survey of the Properties of Artificial Atoms} (North Holland, Amsterdam, 1999);
	L.P. Kouwenhoven, D.G. Austing, and S. Tarucha, Rep. Prog. Phys. {\bf 64}, 701 (2001).
\bibitem{addition_energy}
	S. Tarucha, D.G. Austing, T. Honda, R.J. van der Hage, and L.P. Kouwenhoven, \prl {\bf 77}, 3613 (1996); 
	S.M. Reimann and M. Manninen, Rev. Mod. Phys. {\bf 74}, 1283 (2002).
\bibitem{filinov_2001}
	A.V. Filinov, M. Bonitz, and Yu.E. Lozovik, Phys. Rev. Lett. {\bf 86}, 3851 (2001).
\bibitem{pederiva_2000}
	F. Pederiva, C.J. Umrigar, and E. Lipparini, Phys. Rev. B {\bf 62}, 8120 (2000); {\it ibid.} {\bf 68}, 089901 (2003).
\bibitem{dreizler_book}
	See for instance R.M. Dreizler and E.K.U. Gross, {\it Density Functional Theory, An Approach to the Quantum 
	Many-Body Problem} (Springer, Berlin, 1990).
\bibitem{yannouleas_1999}
	C. Yannouleas and U. Landman, Phys. Rev. Lett. {\bf 82}, 5325 (1999).
\bibitem{perdew_1975}
	D.C. Langreth and J.P. Perdew, Solid State Commun. {\bf 17}, 1425 (1975); 
	O Gunnarsson and B.I. Lundqvist, Phys. Rev. B {\bf 13}, 4274 (1976).
\bibitem{gunnarsson_1989}
	See {\it e.g.} R.O. Jones and O. Gunnarsson, Rev. Mod. Phys. {\bf 61}, 689 (1989).
\bibitem{attaccalite_2002}
	C. Attaccalite, S. Moroni, P. Gori-Giorgi, and G. Bachelet, Phys. Rev. Lett. {\bf 88}, 256601 (2002).
\bibitem{gori_giorgi_2004}
	P. Gori-Giorgi, S. Moroni, and G.B. Bachelet, Phys. Rev. B {\bf 70}, 115102 (2004). 
	These expressions for the PDF of the $2D$ EG are fully consistent with the xc energy of 
	Ref.~\onlinecite{attaccalite_2002}.
\bibitem{abram}
	M. Abramowitz and  I.A. Stegun, {\it Handbook of Mathematical Functions} (Dover, New York, 1972).
\bibitem{szabo_1989}
	A. Szabo and N.S. Ostlund, {\it Modern Quantum Chemistry} (MacGraw-Hill, New York, 1989). 
\bibitem{cohl_1999}
	H.S. Cohl and J.E. Tohline, Astrophys. J. {\bf 527}, 86 (1999). In this work the expansion 
	\[
	\frac{1}{|{\bf r}-{\bf r}'|}=\frac{1}{\pi\sqrt{rr'}}\sum_{m=-\infty}^{m=+\infty}{\mathcal 
	Q}_{m-1/2}(\chi)\exp{[im(\phi-\phi')]}
	\]
	is introduced where ${\bf r}=(r,\phi)$ and $\chi=({\bf r}^2+{\bf r}'^2)/(2rr')$. Here ${\mathcal Q}_{m-1/2}(x)$ is 
	the Legendre function of the second kind with odd-half-integer degree~\cite{abram}.
\bibitem{tanatar_1989}
	B. Tanatar and D.M. Ceperley, Phys. Rev. B {\bf 39}, 5005 (1989).
\bibitem{pdft}
	P. Ziesche, Phys. Lett. A {\bf 195}, 213 (1994); 
	A. Gonis, T.C. Schulthess, J. van Ek, and P.E.A. Turchi, Phys. Rev. Lett. {\bf 77}, 2981 (1996); 
	A. Gonis, T.C. Schulthess, P.E.A. Turchi, and J. van Ek, Phys. Rev. B {\bf 56}, 9335 (1997); 
	M. Levy and P. Ziesche, J. Chem. Phys. {\bf 115}, 9110 (2001); 
	A. Nagy, Phys. Rev. A {\bf 66}, 022505 (2002); 
	F. Furche, Phys. Rev. A {\bf 70}, 022514 (2004).
\bibitem{davoudi_2002}
	B. Davoudi, M. Polini, R. Asgari, and M.P. Tosi, Phys. Rev. B {\bf 66}, 075110 (2002).
\bibitem{overhauser_1995}
	A.W. Overhauser, Can. J. Phys. {\bf 73}, 683 (1995).
\bibitem{gori_giorgi&savin_2004}
	P. Gori-Giorgi and A. Savin, cond-mat/0411179.
\bibitem{ebner_1976}
	C. Ebner, W.F. Saam, and D. Stroud, Phys. Rev. A {\bf 14}, 2264 (1976); 
	W.F. Saam and C. Ebner, {\it ibid}. {\bf 15}, 2566 (1977).
\bibitem{yamashita_1984}
	I. Yamashita and S. Ichimaru, Phys. Rev. B {\bf 29}, 673 (1984).
\bibitem{egger_1999}
	R. Egger, W. H\"{a}usler, C.H. Mak, and H. Grabert, Phys. Rev. Lett. {\bf 82}, 3320 (1999).
\bibitem{reimann_2000}
	S.M. Reimann, M. Koskinen, and M. Manninen, Phys. Rev. B {\bf 62}, 8108 (2000). 
	These authors reported both exact diagonalization and SDFT. Relative to the latter our calculations have used 
	updated input for the xc energy and have been extended to larger vaues of $\lambda$.
\bibitem{ring_1980}
	P. Ring and P. Schuck, {\it The Nuclear Many-Body Problem} (Springer, New York, 1980), pag. 438~ff.
\end{thebibliography}
\end{document}